%% file: paper_main.tex
\documentclass[pra,twocolumn,superscriptaddress,nofootinbib]{revtex4-1}

\usepackage{lipsum}
\usepackage{braket}
\usepackage{graphicx}
\usepackage{dcolumn}
\usepackage{enumitem}
\usepackage{bm}
\usepackage{amsmath}
\usepackage{amssymb}
\usepackage{mathtools}
\usepackage{subfigure}
\usepackage{color}
\usepackage{xcolor}
\usepackage{verbatim}
\usepackage{float}
\usepackage{appendix}
\usepackage{bbold}

\newcommand{\abs}[1]{\lvert #1 \rvert} 
\newcommand{\ev}[1]{\langle #1 \rangle} 
\newcommand{\lv}[0]{\mathcal{L}} 
\newcommand{\td}[1]{\tilde{#1}} 
\newcommand{\nbar}[0]{\bar{n}} 
\newcommand{\mbf}[1]{\mathbf{#1}} 
\newcommand{\hSig}[0]{\hat{\sigma}} 
\newcommand{\mh}[0]{\mathcal{H}} 
\newcommand{\mv}[0]{\mathcal{V}} 
\newcommand{\mt}[0]{\mathcal{T}} 
\newcommand{\exb}[0]{\mathbf{E}\times\mathbf{B}} 

\begin{document}

\title{Broadening of the drumhead mode spectrum due to in-plane thermal fluctuations of two-dimensional trapped ion crystals in a Penning trap}
\date{\today}
\author{Athreya Shankar}
\email{athreya.shankar@colorado.edu}
\affiliation{JILA, NIST and Department of Physics, University of Colorado Boulder}
\author{Chen Tang}
\affiliation{Department of Physics, University of Colorado Boulder}
\author{Matthew Affolter}
\affiliation{Time and Frequency Division, National Institute of Standards and Technology Boulder}
\author{Kevin Gilmore}
\affiliation{Time and Frequency Division, National Institute of Standards and Technology Boulder}
\affiliation{Department of Physics, University of Colorado Boulder}
\author{Daniel H. E. Dubin}
\affiliation{Department of Physics, University of California San Diego}
\author{Scott Parker}
\affiliation{Department of Physics, University of Colorado Boulder}
\author{Murray J. Holland}
\affiliation{JILA, NIST and Department of Physics, University of Colorado Boulder}
\author{John J. Bollinger}
\affiliation{Time and Frequency Division, National Institute of Standards and Technology Boulder}

\begin{abstract}
Two-dimensional crystals of ions stored in Penning traps are a leading platform for quantum simulation and sensing experiments. For small amplitudes, the out-of-plane motion of such crystals can be described by a discrete set of normal modes called the drumhead modes, which can be used to implement a range of quantum information protocols. However, experimental observations of crystals with Doppler-cooled and even near-ground-state-cooled drumhead modes reveal an unresolved drumhead mode spectrum. In this work, we establish in-plane thermal fluctuations in ion positions as a major contributor to the broadening of the drumhead mode spectrum. In the process, we  demonstrate how the confining magnetic field leads to unconventional in-plane normal modes, whose average potential and kinetic energies are not equal. This property, in turn, has implications for the sampling procedure required to choose the in-plane initial conditions for molecular dynamics simulations. For current operating conditions of the NIST Penning trap, our study suggests that the two dimensional crystals produced in this trap undergo in-plane potential energy fluctuations of the order of $10$ mK. Our study therefore motivates the need for designing improved techniques to cool the in-plane degrees of freedom.  
\end{abstract}

\maketitle

\section{Introduction}

In the past decade, large crystals of ions stored in Penning traps have shown great promise for quantum simulation and quantum sensing experiments. Such systems have enabled the study of iconic and important spin and spin-boson models \cite{britton2012nat,safavi2018prl,cohn2018njp}, of the growth of entanglement in large interacting systems \cite{garttner2017nat,swingle2016PRA}, and of metrologically relevant protocols aimed at spin squeezing \cite{bohnet2016sci,pezze2018RMP} and motion sensing \cite{gilmore2017PRL,toscano2006PRA}. A number of efforts are currently underway to expand the toolbox for quantum information processing with Penning traps, including the development of miniaturized permanent-magnet systems that offer portability \cite{mcmahon2020PRA} and traps with improved optical access \cite{ball2019RSI}, the incorporation of sideband cooling \cite{hrmo2019PRA}, and proposals for quantum computing and simulation in arrays of Penning traps \cite{jain2018arXiv}. 

Equilibrium trapped-ion crystals result from a balance between the external trapping fields and the inter-ion Coulomb repulsion. In Penning traps, the trapping fields consist of (a) an electric quadrupole field that ensures confinement along the trap axis ($z$ axis), and (b) a strong magnetic field pointing along the trap axis, that, combined with the electric field, provides radial confinement \cite{dubin1999RMP,biercuk2009QIC}. As a result of the magnetic field, the equilibrium crystal is rotating about the trap axis in the lab frame. This rotation frequency $\omega_r$ can be stabilized by the application of a weak quadrupole rotating wall potential \cite{huang1997PRL,huang1998POP,dubin1999RMP}. In general, Penning traps can be configured to realize three-dimensional crystals. However, in the special case that the radial confinement is weak compared to that along the axial direction, the ions organize into a two-dimensional crystal in the $x-y$ plane \cite{wang2013PRA}. 

These two-dimensional crystals have attractive features for quantum simulation and sensing. First, single-ion laser addressing and detection is simplified in a planar configuration. Second, for small displacements, the out-of-plane and in-plane motions decouple, leading to a simple description of the out-of-plane dynamics in terms of collective normal modes that obey simple harmonic motion with discrete frequencies \cite{wang2013PRA}. These modes are referred to as the drumhead modes and are used to realize the ion-ion interactions that are necessary for quantum simulation and sensing protocols. On the other hand, the nature of the in-plane modes is complicated by the velocity-dependent Lorentz force produced by the magnetic field. Broadly, the Lorentz force splits the in-plane modes into two branches, one consisting of high frequency cyclotron modes and the other consisting of low frequency $\exb$ modes \cite{wang2013PRA}.

\begin{figure*}[!tb]
    \centering
    \includegraphics[width=0.8\textwidth]{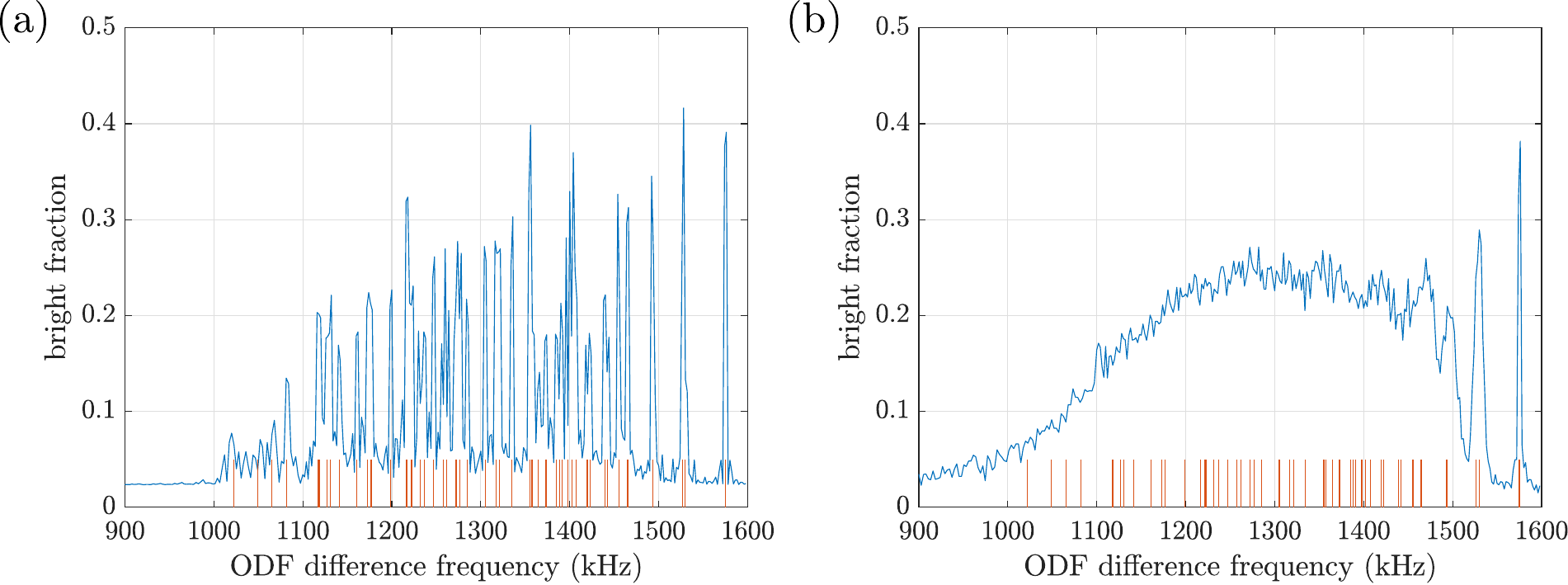}
    \caption{Comparison of (a) theoretically computed and (b) experimentally measured ODF spectra for a Doppler-cooled crystal with $53$ ions. The orange lines indicate the position of the theoretically computed normal modes. The theory computation in (a) is performed according to the procedure described in Sec.~\ref{sec:odf} and assumes the same parameters as that used to experimentally measure the data shown in (b). For the simulation in panel (a), we assume that the ions have zero temperature in the crystal plane, while a temperature of $0.5$ mK is assumed for their out-of-plane motion.}
    \label{fig:n53_odf_paper}
\end{figure*}

Measurements of the motional modes of collections of charged particles that form ion crystals or, more generally, non-neutral plasmas can provide useful diagnostic information on temperatures and densities \cite{shiga2006POP,anderegg2003POP}. Experimentally, the drumhead mode spectrum is probed by the application of an optical dipole force (ODF) that maps the out-of-plane motion at a particular frequency onto the electronic population in a hyperfine level of the ion \cite{sawyer2012PRL,sawyer2012PRA,jordan2019PRL,shankar2020Thesis}. The population in this level can be read out via fluorescence detection and hence this level is referred to as the bright state. The motional frequency that is probed can be varied by adjusting the difference frequency of the pair of lasers used to generate the ODF. The fraction of ions found in the bright state as the ODF difference frequency is scanned over the bandwidth of drumhead modes provides qualitative and quantitative information on the drumhead mode spectrum and is frequently called the ODF spectrum. 

Although linearized theory predicts a discrete set of drumhead mode frequencies, features of experimentally measured ODF spectra are in contradiction with this expectation \cite{sawyer2012PRL,jordan2019PRL}.  Figure~\ref{fig:n53_odf_paper} illustrates the disparity between theory predictions and experimental observations of the ODF spectrum for a crystal with $N=53$ ions that has been subjected to Doppler laser cooling. The theory curve in Fig.~\ref{fig:n53_odf_paper}(a) is computed assuming that the ions undergo thermal motion out-of-plane but have zero temperature in the crystal plane. In this case, the out-of-plane motion is seen to be well described by the linearized theory and the ODF diagnostic partially resolves the fixed, discrete set of drumhead modes over their entire bandwidth, leading to a jagged spectrum. 

However, the experimental data in Fig.~\ref{fig:n53_odf_paper}(b) reveals that only the first few highest frequency drumhead modes are well resolved. In particular, the  locations of the three highest frequency peaks in the experimental data are well described by the normal mode theory. The highest peak marks the axial center of mass mode, whose frequency is independent of the particle positions in an ideal harmonic trap and is given by the axial trapping frequency $\omega_\parallel$ (cf. Table~\ref{tab:char_freq} and Eq.~(\ref{eqn:eff_pot_sim})). The next two peaks respectively correspond to tilt modes, i.e. tilts of the lattice out of the crystal plane, and `potato chip' modes, which are long-wavelength bends to the lattice \cite{wang2013PRA}. A cold fluid theory for these modes \cite{dubinPRL1991,wiemer1994PRA} predicts frequencies of  $\sqrt{\omega_\parallel^2-\omega_\perp^2}$ and  $\sqrt{\omega_\parallel^2- 7\omega_\perp^2/4}$ respectively, where $\omega_\perp$ is the effective radial trapping frequency (cf. Table~\ref{tab:char_freq} and Eq.~(\ref{eqn:eff_pot_sim})). For large $N$ the wavelengths of these modes are large compared to the inter-ion spacing and therefore the frequencies of these modes are not very sensitive to the underlying crystal configuration. Thus, the widths of these peaks are small and their locations are in good agreement with predictions from the cold fluid theory. However, below these modes, the measured spectrum is essentially a smooth continuum over frequencies, extending approximately until the lowest predicted mode frequency. In Ref.~\cite{jordan2019PRL}, we attempted to numerically reproduce qualitative features of the observed spectrum by incorporating \emph{ad hoc} mode frequency fluctuations, assuming that the variance increases with decreasing mode frequency. 

The objective of this paper is to develop a first-principles model to understand and explain the broadening of the drumhead mode spectrum. Our study strongly suggests that low frequency in-plane thermal fluctuations in ion positions, attributable to the $\exb$ modes, are an important contributor to the broadening of the drumhead spectrum. 

Such a physical model for mode frequency fluctuations is an essential starting point for attempts at improving the mode resolution and in identifying opportunities for refining the experimental system. First, improved mode resolution will greatly expand the range of simulation and sensing protocols possible with Penning traps; so far, ion-ion interactions have been primarily mediated by the well-resolved center-of-mass (c.m.) mode, which is also the highest frequency drumhead mode. A variety of network topologies can be realized if a large number of drumhead modes are well resolved and individually addressable like the c.m. mode \cite{shapira2020PRA}. Second, for parameters of the current experiments with the NIST Penning trap \cite{jordan2019PRL}, we typically find that in-plane fluctuations in the ion crystal potential energy of the order of $10$ mK are required to numerically simulate drumhead spectra as poorly resolved as the experimental measurements, indicating the need for improved cooling of the in-plane degrees of freedom.

This paper is organized as follows. In Section~\ref{sec:prelim}, we first summarize the model for the dynamics of two-dimensional trapped ion crystals in a Penning trap. Next, we linearize the equations of motion and introduce the formalism to study the in-plane normal modes. In Section~\ref{sec:ratio}, we illustrate the unconventional nature of the in-plane modes by investigating the ratio $R_n$ of the average potential energy to the average kinetic energy in mode $n$. In the process, we show that the in-plane modes can be attributed a direction of rotation because of the presence of the magnetic field. In Section~\ref{sec:insights}, we discuss the implications of $R_n \neq 1$ for the in-plane thermal fluctuations and for the initialization procedure to be used for thermal molecular dynamics simulations of the crystal motion. We then proceed to discuss drumhead mode frequency fluctuations in Section~\ref{sec:drumhead}. We first present an intuitive picture for the origin of these fluctuations that is based on static thermal snapshots of in-plane crystal configurations. Along the lines of the Shannon entropy, we introduce a mode entropy to quantify the extent of delocalization of a mode, by which we are able to qualitatively explain the varying sensitivity of different drumhead modes to in-plane thermal fluctuations. We then validate the findings from this intuitive picture by performing molecular dynamics simulations with appropriate initialization of the in-plane coordinates. We use these simulations to study the power spectral density of the drumhead motion in the presence of in-plane thermal fluctuations. We also extract ODF spectra from these simulations, which can be used for direct qualitative comparisons to experimental measurements. Finally, we summarize our findings and conclude with an outlook in Section~\ref{sec:conc}.

\section{\label{sec:prelim}Preliminaries}

We consider a collection of $N$ ions of mass $m$ and charge $e$ in a Penning trap. The magnetic field pointing along the $z$ axis has strength $B_z>0$, while the electric quadrupole potential is parametrized by an amplitude $V_0>0$. The strength of the magnetic field can also be quantified by the bare cyloctron frequency $\omega_c = eB_z/m$. In addition, we assume that a rotating wall potential with amplitude $V_W>0$ and frequency $\omega_r>0$ is applied to stabilize the crystal rotation frequency. A crystal stored in this trap rigidly rotates with angular frequency $\omega_r$ in the clockwise sense when viewed from above the crystal plane in the lab frame, i.e. when the line-of-sight is along the $-z$ axis. In a frame rotating at angular frequency $\omega_r$, the Lagrangian becomes time independent and is given by \cite{wang2013PRA}
\begin{eqnarray}
    \lv &=& \sum_{j=1}^N \left[\frac{m}{2} \dot{\mbf{r}}_j \cdot \dot{\mbf{r}}_j -e\phi_j -\frac{m \omega_c'}{2}(\dot{x}_j y_j - \dot{y}_j x_j) \right],
    \label{eqn:lag_rot}
\end{eqnarray}
where $\omega_c' = \omega_c-2\omega_r$ is an effective cyclotron frequency and 
\begin{eqnarray}
    e\phi_j &=& eV_0 z_j^2 +\frac{1}{2}\left(eB_z\omega_r-m\omega_r^2-eV_0\right)(x_j^2+y_j^2) \nonumber\\
    &&+ eV_W (x_j^2-y_j^2) + \frac{k_e e^2}{2}\sum_{k\neq j}\frac{1}{r_{kj}}
    \label{eqn:eff_pot}
\end{eqnarray}
is the effective potential energy of ion $j$. Here, $x_j,y_j$ are rotating frame coordinates and are related to the lab frame coordinates through a time dependent rotation matrix \cite{wang2013PRA}. The position of an ion in the rotating frame is compactly denoted by the vector $\mbf{r}_j = x_j \mbf{\hat{x}} + y_j \mbf{\hat{y}} + z_j \mbf{\hat{z}}$. Finally, the Coulomb potential is expressed using $r_{kj}=\sqrt{(x_k-x_j)^2+(y_k-y_j)^2+(z_k-z_j)^2}$ and  $k_e=1/(4\pi\epsilon_0)$, where $\epsilon_0$ is the permittivity of free space.

\begin{table}[!tb]
\begin{tabular}{cc}
Frequency     &   Description \\\hline
$\omega_c = eB_z/m$ & bare cyclotron \\
$\omega_c'=\omega_c-2\omega_r$  & effective cyclotron \\
$\omega_\parallel = \sqrt{2eV_0/m}$  & axial trapping \\
$\omega_\perp = \sqrt{\omega_c\omega_r-\omega_r^2-\omega_\parallel^2/2}$ & radial trapping \\
$\omega_W = \sqrt{2eV_W/m}$ & wall correction 
\end{tabular}
\caption{Some characteristic frequencies governing the dynamics of ions in a frame rotating at frequency $\omega_r$ in a Penning trap. With the exception of Fig.~\ref{fig:n53_odf_paper}, the numerical results presented in this paper correspond to a $120$ ion crystal in a trap characterized by the frequencies  $\omega_c/2\pi \approx 7.60$ MHz, $\omega_r/2\pi=180$ kHz, $\omega_\parallel/2\pi=1.59$ MHz, and $\omega_W/2\pi \approx  68$ kHz. These values are typical of the current operational parameters for producing crystals of ${}^9\text{Be}^+$ ions in the NIST Penning trap.}
\label{tab:char_freq}
\end{table}

Table~\ref{tab:char_freq} summarizes some characteristic frequencies in terms of which we rewrite the Lagrangian, Eq.~(\ref{eqn:lag_rot}). Specifically, the effective potential energy in Eq.~(\ref{eqn:eff_pot}) simplifies to 
\begin{eqnarray}
    e\phi_j &=& \frac{m\omega_\parallel^2}{2}z_j^2 
    + \frac{m\omega_\perp^2}{2}(x_j^2+y_j^2) \nonumber\\
    &&+ \frac{m\omega_W^2}{2}(x_j^2-y_j^2)
    + \frac{k_e e^2}{2}\sum_{k\neq j}\frac{1}{r_{kj}}.
    \label{eqn:eff_pot_sim}
\end{eqnarray}
In this paper, we use the $\parallel$ ($\perp$) subscript to indicate motion parallel (perpendicular) to the direction of the magnetic field.

\subsection{Euler-Lagrange equations of motion}
We obtain the equations of motion along the three directions as a set of first order coupled differential equations in the positions and velocities. The positions are updated trivially as 
\begin{equation}
    \dot{\mbf{r}}_j = \mbf{v}_j
    \label{eqn:pos_update}
\end{equation}
by construction. The velocity components evolve as 
\begin{eqnarray}
    \dot{v}_j^x &=& -\left(\omega_\perp^2+\omega_W^2\right)x_j + \omega_c' v_j^y + \frac{k_e e^2}{m}\sum_{k \neq j}\frac{x_j-x_k}{r_{kj}^3}, \nonumber\\
    \dot{v}_j^y &=& -\left(\omega_\perp^2-\omega_W^2\right)y_j - \omega_c' v_j^x + \frac{k_e e^2}{m}\sum_{k \neq j}\frac{y_j-y_k}{r_{kj}^3}, \nonumber\\
    \dot{v}_j^z &=& -\omega_\parallel^2 z_j + \frac{k_e e^2}{m}\sum_{k \neq j}\frac{z_j-z_k}{r_{kj}^3}.
    \label{eqn:mom_update}
\end{eqnarray}

\subsection{Linearized equations about stable equilibrium}

A stable equilibrium two-dimensional crystal can be found by determining the configuration of ions in the $x-y$ plane that minimizes the total potential energy given by $\Phi = \sum_{j=1}^N e\phi_j$. Numerically, we perform the minimization by starting from a seed lattice and iteratively finding a local minimum, as described in Ref.~\cite{wang2013PRA}. Alternatively, a global minimization technique such as simulated annealing can also be employed which eliminates the need for an initial guess (seed) configuration \cite{shankar2020Thesis}.

We denote the equilibrium ion coordinates of ion $j$ by $x_j^0,y_j^0,z_j^0$. We can then write $s_j =  s_j^0 + \delta s_j$ for each coordinate $s\rightarrow x,y,z$ and consider the situation where the displacements $\delta s_j$ are small compared to the lattice spacing. In this limit, we can approximate  
\begin{equation}
    r_{kj}^{-3} \approx \left(\left(r_{kj}^0\right)^2 + 2\sum_{s\rightarrow x,y}(s_j^0-s_k^0)(\delta s_j-\delta s_k)\right)^{-3/2}.
    \label{eqn:rkj_approx}
\end{equation}
Since $z_j^0=0 \; \forall \; j$ in the case of 2D crystals, $r_{kj}^0 = \sqrt{(x_k^0-x_j^0)^2+(y_k^0-y_j^0)^2}$ and the $s \rightarrow z$ term is omitted in the summation appearing in Eq.~(\ref{eqn:rkj_approx}). Taylor expanding Eq.~(\ref{eqn:rkj_approx}) gives 
\begin{equation}
    \frac{1}{r_{kj}^3} \approx \frac{1}{\left(r_{kj}^0\right)^3} -3\sum_{s\rightarrow x,y}\frac{(s_j^0-s_k^0)(\delta s_j-\delta s_k)}{\left(r_{kj}^0\right)^5}.
    \label{eqn:rkj_taylor}
\end{equation}
We now make the substitution $s_j\rightarrow s_j^0 + \delta s_j$ in Eq.~(\ref{eqn:mom_update}) and use the result Eq.~(\ref{eqn:rkj_taylor}). At zeroth order in $\delta s$, the RHS of the velocity update equations vanish because the force is zero in the equilibrium configuration. At first order in $\delta s$, the equations can be expressed in matrix notation as 
\begin{eqnarray}
    \frac{d}{dt}\ket{v_\perp} &=& -\frac{\mathbb{K}_\perp}{m} \ket{\delta r_\perp} + \mathbb{L} \ket{v_\perp}, \nonumber\\
    \frac{d}{dt}\ket{v_\parallel} &=& -\frac{\mathbb{K}_\parallel}{m} \ket{\delta r_\parallel}.
    \label{eqn:linear_eom}
\end{eqnarray}
The vectors $\ket{v_\perp},\ket{\delta r_\perp}$ ($\ket{v_\parallel},\ket{\delta r_\parallel}$) are $2N$ ($N$) dimensional as they account for the $x,y$ ($z$) degrees of freedom of all the ions. Here, $\mathbb{K}_\perp$ and $\mathbb{K}_\parallel$ are real, symmetric stiffness matrices whose forms, for example, can be found in Ref.~\cite{wang2013PRA}. Since they describe motion about a stable equilibrium, both stiffness matrices are positive definite. In particular, a non-zero rotating wall strength, i.e. $\abs{V_W}>0$, breaks the cylindrical symmetry about the trap axis ensuring that the eigenvalues of $\mathbb{K}_\perp$ are strictly non-zero. The out-of-plane and in-plane dynamics decouple in the linear analysis. From the second line of Eq.~(\ref{eqn:linear_eom}), the out-of-plane motion can clearly be described by a set of decoupled simple harmonic oscillators. Following standard normal mode analysis \cite{goldstein2011classical}, the discrete frequency components of the out-of-plane motion can be determined from the eigenvalues of $\mathbb{K}_\parallel$. However, the in-plane dynamics is complicated by the Lorentz force entering through the antisymmetric, block diagonal matrix $\mathbb{L}$ that will be discussed in more detail in Section~\ref{sec:hel}.

\subsection{Hamiltonian}

We write down the Hamiltonian corresponding to Eq.~(\ref{eqn:lag_rot}) using the position and velocity coordinates, instead of the canonical momenta. In these coordinates, the Hamiltonian is separable and can be written as
\begin{equation}
    \mh = \mt + \mv,
\end{equation}
where $\mt=m\left(\braket{v_\perp | v_\perp} +\braket{v_\parallel | v_\parallel}\right)/2 $ is the kinetic energy and $\mv=\sum_{j=1}^N e\phi_j$ is the potential energy.  
In the harmonic approximation, $\mv$ can be expressed as 
\begin{equation}
    \mv = \mv_0 + \frac{1}{2} \braket{ \delta r_\perp | \mathbb{K}_\perp | \delta r_\perp} + \frac{1}{2} \braket{ \delta r_\parallel |  \mathbb{K}_\parallel | \delta r_\parallel},   
\end{equation}
where $\mv_0$ is the potential energy of the equilibrium configuration. The Hamiltonians for the in-plane and out-of-plane dynamics also decouple in the harmonic approximation. We can identify the excess in-plane energy arising from position and velocity fluctuations as  
\begin{equation}
    \mh_\perp = \mt_\perp + \mv_\perp =\frac{m}{2}\braket{v_\perp | v_\perp}
    +\frac{1}{2}\braket{\delta r_\perp | \mathbb{K}_\perp | \delta r_\perp}.
\end{equation}

\subsection{Composite phase space vector}
By introducing a composite phase space vector
\begin{equation}
    \ket{q_\perp} = \begin{pmatrix}
                    \ket{\delta r_\perp} \\
                    \ket{v_\perp}
                    \end{pmatrix},
\end{equation}
the in-plane dynamics in the harmonic approximation can be expressed as $d\ket{q_\perp}/dt = \mathbb{D} \ket{q_\perp}$ \cite{dubin2020arXiv}, where $\mathbb{D}$ is a $4N \times 4N$ matrix given by 
\begin{equation}
    \mathbb{D} = \begin{pmatrix}
        \mathbb{0}_{2N\times2N} & \mathbb{1}_{2N} \\
        -\frac{\mathbb{K}_\perp}{m} & \mathbb{L}
        \end{pmatrix}.
    \label{eqn:dmat}    
\end{equation}
In writing Eq.~\ref{eqn:dmat}, we have introduced the notations $\mathbb{0}_{r \times c}$ and $\mathbb{1}_{d}$ to respectively express a $r\times c$ dimensional zero matrix and the $d \times d$ dimensional identity matrix. The in-plane Hamiltonian can similarly be compactly expressed as $\mh_\perp = \braket{q_\perp | \mathbb{E} | q_\perp}/2$, where  
\begin{equation}
\mathbb{E}=\text{diag}\left(\mathbb{K}_\perp,m\mathbb{1}_{2N}\right)    
\end{equation} 
is an energy matrix. We use the notation $\allowdisplaybreaks\text{diag}\left(\mathbb{M}_1,\mathbb{M}_2,\ldots,\mathbb{M}_N\right)$ to represent the block diagonal matrix obtained as the direct sum of the matrices $\mathbb{M}_1,\mathbb{M}_2,\ldots,\mathbb{M}_N$.

\subsection{Generalized eigenvalue problem}

We make the ansatz $\ket{q_\perp} \sim \ket{u_n}e^{-i\omega_n t}$, where $n=1,\ldots 4N$ labels the real eigenvalues $\omega_n$. The vectors $\ket{u_n}$ satisfy the eigenvalue equation 
\begin{equation}
    \mathbb{D} \ket{u_n} = -i\omega_n \ket{u_n}.
    \label{eqn:eig_prob}
\end{equation}
Since $\mathbb{D},\omega_n$ are real, complex conjugation of Eq.~(\ref{eqn:eig_prob}) reveals that $\ket{u_n^*}$ is also an eigenvector with eigenvalue $-\omega_n$. Therefore, we can choose the first $2N$ eigenvectors to have $\omega_n>0$, while the last $2N$ eigenvectors are simply related as $\omega_{2N+n}=-\omega_n$, for $n=1,\ldots,2N$.

Multiplying Eq.~(\ref{eqn:eig_prob}) by $\mathbb{E}$, we get 
\begin{equation}
    \mathbb{D}_H \ket{u_n} = \omega_n \mathbb{E}\ket{u_n},
    \label{eqn:gep}
\end{equation}
where the Hermitian matrix $\mathbb{D}_H$ is 
\begin{equation}
    \mathbb{D}_H = i\begin{pmatrix}
                    0   &   \mathbb{K}_\perp \\
                    -\mathbb{K}_\perp  &   m\mathbb{L}
                    \end{pmatrix}.
\end{equation}
Equation~(\ref{eqn:gep}) is a generalized eigenvalue problem of the form $\mathbb{A}\ket{u_n}=\omega_n\mathbb{B}\ket{u_n}$ with Hermitian matrices $\mathbb{A}$ and $\mathbb{B}$. Therefore, the eigenvectors are $\mathbb{E}$-orthogonal, i.e. for $k \neq n$, they satisfy 
\begin{equation}
    \braket{u_k | \mathbb{E} | u_n} = 0.
    \label{eqn:e_ortho}
\end{equation}

In terms of these eigenvectors, the phase space vector can be written as 
\begin{equation}
    \ket{q_\perp} = \sum_{n=1}^{2N} \left(A_n e^{-i\omega_n t}  \ket{u_n} + A_n^* e^{i\omega_n t}\ket{u_n^*}\right),
    \label{eqn:q_harmonic}
\end{equation}
where $\omega_{2N+n}=-\omega_n$ and $\ket{u_{2N+n}}=\ket{u_n^*}$ have been used, the $A_n$ are complex initial amplitudes for each mode and $A_{2N+n}=A_n^*$ has been imposed to ensure that $\ket{q_\perp}$ is real. 

\subsection{Potential and kinetic energy in a mode}

Using Eq.~(\ref{eqn:q_harmonic}) in the expression $\mh_\perp = \braket{q_\perp | \mathbb{E} | q_\perp}/2$, the in-plane energy can be expanded in terms of a double sum over the normal modes. Using the orthogonality relation, Eq.~(\ref{eqn:e_ortho}), the expression can be reduced to a single sum over the $2N$ in-plane modes given by 
\begin{eqnarray}
    \mh_\perp &=& \sum_{n=1}^{2N} \abs{A_n}^2 \braket{u_n | \mathbb{E} | u_n} \nonumber\\
    &=& \sum_{n=1}^{2N} \abs{A_n}^2  \left(\braket{u_n^r | \mathbb{K}_\perp | u_n^r} +  m \braket{u_n^v | u_n^v} \right),
    \label{eqn:hperp}
\end{eqnarray}
where we have denoted the position and velocity components of $\ket{u_n}$ respectively by $\ket{u_n^r}$ and $\ket{u_n^v}$. 
The position and velocity contributions to the energy in each mode can respectively be identified as the DC component or the time-average of the potential and kinetic energy in that mode.   

\section{\label{sec:ratio}Ratio of potential and kinetic energies}

In this Section, we investigate a characteristic property of any given normal mode $n$, namely, the ratio $R_n$ of the DC component of the potential and kinetic energies, which is defined as 
\begin{equation}
    R_n = \frac{\braket{u_n^r | \mathbb{K}_\perp | u_n^r}}{m \braket{u_n^v | u_n^v}}.
    \label{eqn:rn_1}
\end{equation}
Clearly, $R_n>0$ since $\mathbb{K}_\perp$ is positive definite. 
Expanding Eq.~(\ref{eqn:eig_prob}), we observe that
\begin{eqnarray}
&&\ket{u_n^v} = -i\omega_n\ket{u_n^r}, \nonumber\\
-&&\mathbb{K}_\perp \ket{u_n^r} + m\mathbb{L}\ket{u_n^v} = -i m\omega_n \ket{u_n^v}.
\label{eqn:eig_expand}
\end{eqnarray}
Using the first equation, we get 
\begin{equation}
    R_n = \frac{\braket{u_n^r | \mathbb{K}_\perp | u_n^r}}{m\omega_n^2 \braket{u_n^r | u_n^r}}.
    \label{eqn:rn_2}
\end{equation}
If $\mathbb{L}=0$, Eqs.~(\ref{eqn:eig_expand}) imply that $\mathbb{K}_\perp \ket{u_n^r} = m\omega_n^2 \ket{u_n^r}$, i.e. the dynamics reduces to that of a simple harmonic oscillator (SHO) and $R_n=1$ as expected.

However, such an equal contribution of potential and kinetic energies is drastically altered when $\mathbb{L}\neq 0$. We can already anticipate a dramatic modification by gaining intuition from the in-plane normal modes of a single trapped ion.

\subsection{Single ion case}

We substitute the ansatz $\ket{q_\perp} = e^{-i\omega_n t}\ket{q_0}$ in Eq.~(\ref{eqn:pos_update}) and Eq.~(\ref{eqn:mom_update}) and solve for the positive eigenvalues. We ignore the small correction due to $\omega_W$ in this analysis. The resulting mode frequencies, in the frame rotating at frequency $\omega_r$, are 
\begin{equation}
    \omega_\pm = \frac{\sqrt{\omega_c'^2+4\omega_\perp^2}\pm \omega_c'}{2}.
    \label{eqn:n1_omegapm}
\end{equation}
The Lorentz force arising from the magnetic field, entering as the $\omega_c'$ term in Eq.~(\ref{eqn:mom_update}), splits the two bare SHO modes at $\omega_\perp$ into a high frequency cyclotron mode at $\omega_+$ and a low frequency $\exb$ mode, typically called the magnetron mode, at $\omega_-$ such that $\omega_+\omega_- = \omega_\perp^2$. The hierarchy of frequencies is apparent in the $\omega_\perp \ll \omega_c'$ limit, where we have $\omega_- \ll \omega_\perp \ll \omega_+$. Similarly, the eigenvectors are easily found to be 
\begin{equation}
    \ket{u_\pm^r} = \frac{1}{\sqrt{2}}
                  \begin{pmatrix}
                  1 \\
                  \mp i
                  \end{pmatrix},
\end{equation}
corresponding to clockwise circular motion at $\omega_+$ and counterclockwise circular motion at $\omega_-$. In the single-ion case, $\mathbb{K}_\perp = m\omega_\perp^2 \mathbb{I}_2$ trivially, leading to 
\begin{equation}
    R_+ = \frac{\omega_-}{\omega_+} \ll 1, \;
    R_- = \frac{\omega_+}{\omega_-} \gg 1.
\end{equation}
Therefore, the energy of the cyclotron mode is almost entirely kinetic whereas the energy of the magnetron mode is almost entirely potential.

Intuitively, these results can be understood by considering a particle undergoing uniform circular motion at fixed radius $r$ and angular frequency $\omega_\pm$ in a 2D harmonic trap with trapping frequency $\omega_\perp$. The potential energy is then fixed at $m\omega_\perp^2 r^2/2$, whereas the kinetic energy is mode dependent and is given by $m \omega_\pm^2 r^2/2$. Taking the ratio, we immediately recover $R_+ \ll 1$ and $R_- \gg 1$.

\subsection{Multi-ion case}

\begin{figure}[!tb]
    \centering
    \includegraphics[width=0.9\columnwidth]{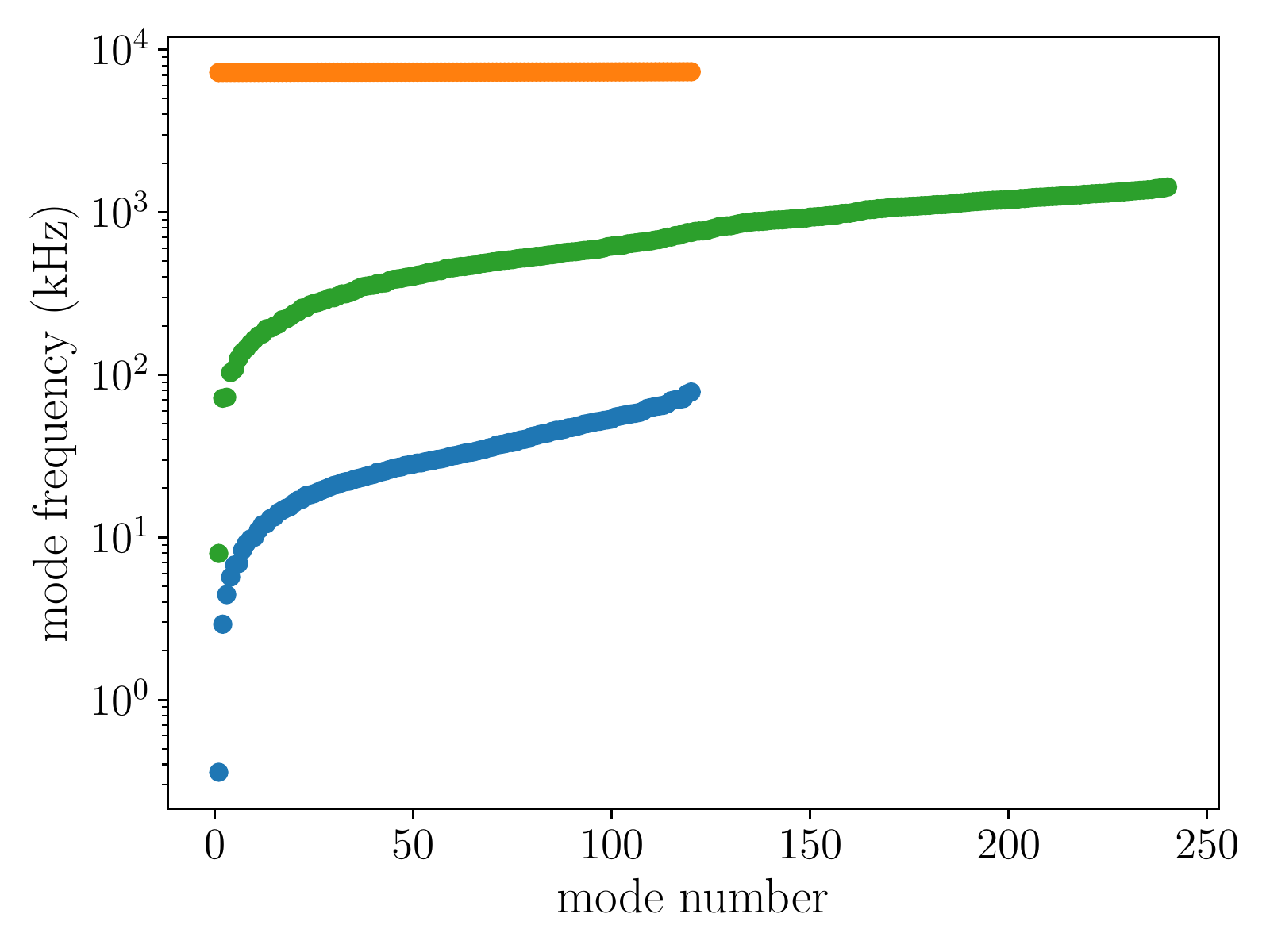}
    \caption{Hierarchy of in-plane mode frequencies. The $2N$ modes (green) corresponding to the bare simple harmonic oscillator system with stiffness matrix $\mathbb{K}_\perp$ are split by the Lorentz force into a low frequency $\exb$ branch (blue) and a high cyclotron branch (orange). Relevant trap and crystal parameters are reported in the caption of Table~\ref{tab:char_freq}.}
    \label{fig:n120_hierarchy}
\end{figure}

In line with the single-ion picture, numerical studies of multi-ion crystals reveal that the $2N$ bare frequencies associated with the stiffness matrix $\mathbb{K}_\perp$ are split by the  Lorentz force, entering through the matrix $\mathbb{L}$ in Eq.~(\ref{eqn:linear_eom}), into $N$ low frequency $\exb$ modes and $N$ high frequency cyclotron modes as shown in Fig.~\ref{fig:n120_hierarchy}. The parameters used correspond to typical operating conditions for the NIST Penning trap. In Section~\ref{sec:rn_chin}, we will use this hierarchy to understand certain features of the $\exb$ and cyclotron branches. 
 
\begin{figure*}[!t]
    \centering
    \includegraphics[width=0.8\textwidth]{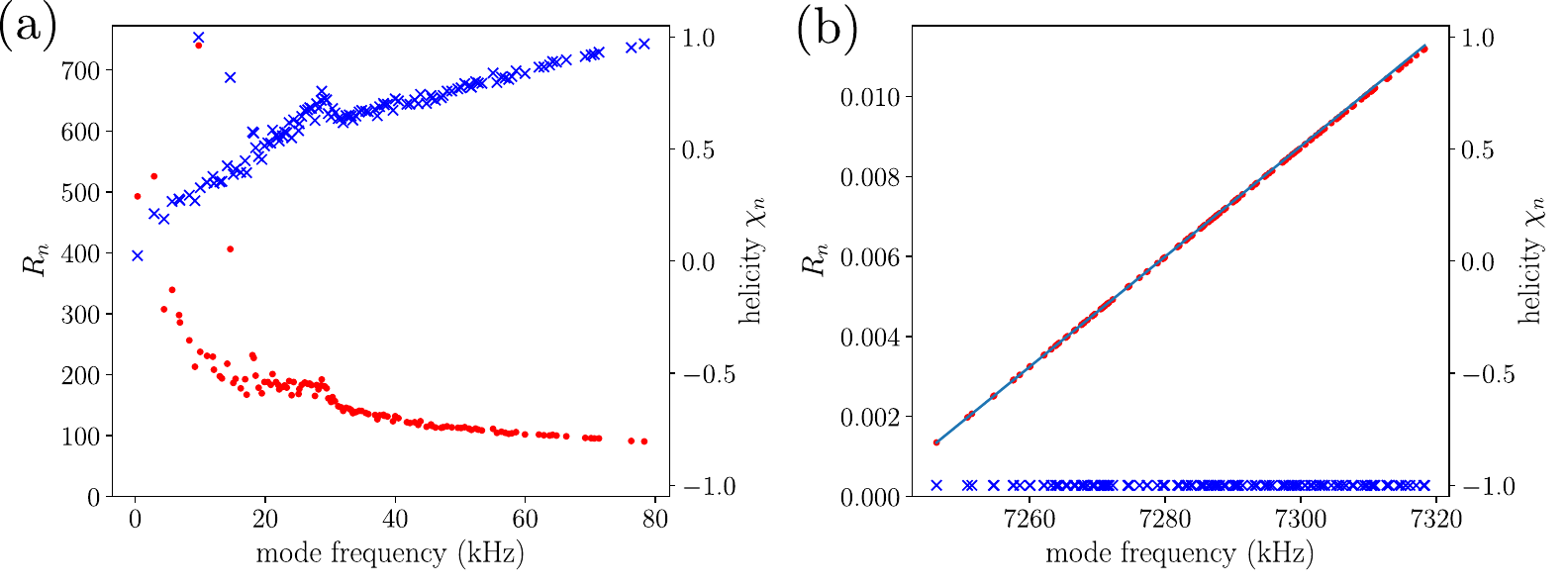}
    \caption{Ratio of potential to kinetic energy $R_n$ (circles) and the helicity $\chi_n$ (crosses) as a function of mode frequency in (a) the $\exb$ branch and (b) the cyclotron branch. The solid line in (b) is an analytic formula, Eq.~(\ref{eqn:rn_cyc}), for the variation of $R_n$ with frequency in the cyclotron branch. Relevant trap and crystal parameters are reported in the caption of Table~\ref{tab:char_freq}.}
    \label{fig:n120_r_chi}
\end{figure*}

As a generalization of the single-ion case, modes in the low frequency branch have $R_n \gg 1$, whereas modes in the high frequency branch display $R_n \ll 1$. Figure~\ref{fig:n120_r_chi} shows the variation of $R_n$ in the two branches. These results indicate that most of the energy in the $\exb$ branch is potential in nature, while the cyclotron modes are dominated by kinetic energy.

\subsubsection{\label{sec:hel}Helicity}

Further insight into the behavior of $R_n$ in the two branches can be obtained by analyzing Eq.~(\ref{eqn:rn_2}) in more detail. From Eq.~(\ref{eqn:eig_expand}), we get 
\begin{equation}
    \braket{u_n^r | \mathbb{K}_\perp | u_n^r} = m\omega_n^2 \braket{u_n^r | u_n^r} + m\omega_n \braket{ u_n^r | (-i\mathbb{L}) | u_n^r}.
    \label{eqn:unr}
\end{equation}
Using this relation in Eq.~(\ref{eqn:rn_2}) gives 
\begin{equation}
    R_n = 1 + \frac{\braket{ u_n^r | (-i\mathbb{L}) | u_n^r}}{\omega_n \braket{u_n^r | u_n^r} }.
    \label{eqn:rn_3}
\end{equation}
The matrix $-i\mathbb{L}$ is Hermitian because $\mathbb{L}$ is antisymmetric. Furthermore, with the basis ordered as $\{x_1,y_1,\ldots,x_N,y_N\}$, $-i\mathbb{L}$ can be expressed as 
\begin{equation}
    \frac{-i\mathbb{L}}{\omega_c'} = \text{diag}\left(\hSig_1^y,\hSig_2^y,\ldots,\hSig_N^y\right),
\end{equation}
where $\hSig_j^y$ is the Pauli $y$ matrix in the basis of $x_j,y_j$. The block diagonal form allows for trivial diagonalization of $-i\mathbb{L}/\omega_c'$, which has eigenvalues $\lambda_{\pm}^j = \mp 1$ and $2N\times 1$ dimensional eigenvectors 
\begin{equation}
    \ket{\lambda_\pm^j} = \begin{pmatrix}
                        \mathbb{0}_{2(j-1)\times1}\\
                        \frac{1}{\sqrt{2}}
                        \begin{pmatrix}
                        1 \\
                        \mp i
                        \end{pmatrix}\\
                        \mathbb{0}_{2(N-j)\times1}
                        \end{pmatrix}.
\end{equation}
The vector $\ket{\lambda_+^j}$ ($\ket{\lambda_-^j}$) has the usual interpretation of clockwise (counterclockwise) circular motion for ion $j$. We expand  $-i\mathbb{L} = \omega_c'\sum_{j=1}^{N} \left(\ket{\lambda_-^j}\bra{\lambda_-^j} -  \ket{\lambda_+^j}\bra{\lambda_+^j}\right)$ and define the helicity $\chi_n$ of a mode as 
\begin{equation}
    \chi_n = \frac{\braket{ u_n^r | (-i\mathbb{L}) | u_n^r}}{\omega_c' \braket{u_n^r | u_n^r}} = \frac{\sum_{j=1}^{N} \left(\abs{\braket{\lambda_-^j | u_n^r}}^2 -  \abs{\braket{\lambda_+^j | u_n^r}}^2\right)}{\sum_{j=1}^{N} \left(\abs{\braket{\lambda_-^j | u_n^r}}^2 +  \abs{\braket{\lambda_+^j | u_n^r}}^2\right)}.
    \label{eqn:chin}
\end{equation}
The helicity of a mode quantifies the normalized tendency to rotate in the counterclockwise ($0<\chi_n\leq1$) or clockwise ($-1\leq \chi_n < 0$) direction. In terms of $\chi_n$, $R_n$ takes the simple form 
\begin{equation}
    R_n = 1 +\frac{\omega_c'}{\omega_n}\chi_n.
    \label{eqn:rn_4}
\end{equation}

Figure~\ref{fig:n120_r_chi} also plots the helicity as a function of mode frequency in the two branches. In the low frequency branch, $\chi_n>0$, indicating that all $\exb$ modes effectively rotate counterclockwise. In contrast, in the high frequency branch, $\chi_n\approx -1$ independent of mode frequency, suggesting that all the cyclotron modes are essentially linear superpositions of pure clockwise rotation of all the ions.

\subsubsection{\label{sec:rn_chin}Mode hierarchy, $R_n$ and $\chi_n$}

The hierarchy of modes, as exemplified in Fig.~\ref{fig:n120_hierarchy}, can be used along with Eq.~(\ref{eqn:unr}) to motivate the observed trends for $R_n$ and $\chi_n$ in the two branches. 

In the $\exb$ branch, the mode frequencies can be assumed small compared to $\omega_c'$ and the bare frequencies associated with $\mathbb{K}_\perp$. Accordingly, Eq.~(\ref{eqn:unr}) can be approximated as 
\begin{equation}
    m\omega_n \braket{ u_n^r | (-i\mathbb{L}) | u_n^r} \approx \braket{u_n^r | \mathbb{K}_\perp | u_n^r}.
\end{equation}
Since $\mathbb{K}_\perp$ is positive definite, we observe that $\braket{ u_n^r | (-i\mathbb{L}) | u_n^r}>0$ within this approximate relation. That is, the helicity $\chi_n>0$ in the $\exb$ branch. Furthermore, since $\omega_n\ll \omega_c'$, we may expect $R_n \approx \chi_n \omega_c'/\omega_n \gg 1$. Both of these trends are indeed observed in Fig.~\ref{fig:n120_r_chi}(a).

In the cyclotron branch, the mode frequencies can be assumed large compared to the bare frequencies associated with $\mathbb{K}_\perp$. Equation~(\ref{eqn:unr}) then simplifies to 
\begin{equation}
     \braket{ u_n^r | (-i\mathbb{L}) | u_n^r} \approx   -\omega_n \braket{u_n^r | u_n^r}.
     \label{eqn:cyc_approx}
\end{equation}
Directly using the relation Eq.~(\ref{eqn:cyc_approx}) in Eq.~(\ref{eqn:rn_3}) gives $R_n \approx 0$. However, we can obtain an improved estimate by first using Eq.~(\ref{eqn:cyc_approx}) in Eq.~(\ref{eqn:chin}), which shows that $\chi_n \approx -1$ in the cyclotron branch. This result is confirmed by the numerical data shown in Fig.~\ref{fig:n120_r_chi}(b). Next, substituting $\chi_n=-1$ in Eq.~(\ref{eqn:rn_4}) gives $R_n \approx 1 -\omega_c'/\omega_n$ in the cyclotron branch. Ignoring the correction due to a rotating wall potential, the frequency of the lowest cyclotron mode in the multi-ion case coincides with $\omega_+$, which is the lone cyclotron mode in the single ion case (see Eq.~(\ref{eqn:n1_omegapm})). Since the bandwidth of the cyclotron modes is small compared to $\omega_+$, we can approximate
\begin{equation}
    \frac{\omega_c'}{\omega_n} \approx \frac{\omega_c'}{\omega_+} - \frac{\omega_c'}{\omega_+^2}\left(\omega_n-\omega_+\right).
\end{equation}
The expression for $R_n$ then reduces to 
\begin{equation}
    R_n \approx 1 - \frac{\omega_c'}{\omega_+} + \frac{\omega_c'}{\omega_+^2}\left(\omega_n-\omega_+\right).
    \label{eqn:rn_cyc}
\end{equation}
Equation~(\ref{eqn:rn_cyc}) explains the linear relation between $R_n$ and $\omega_n$ in the cyclotron branch. Figure~\ref{fig:n120_r_chi}(b) shows the excellent agreement between the prediction from Eq.~(\ref{eqn:rn_cyc}) (solid line) and the numerical data (circles).

\section{\label{sec:insights}Insights enabled by $R_n$}

The ratio $R_n$ is a useful property to better understand the in-plane modes since it enables us to quantify deviation from simple harmonic motion ($R=1$). In this Section, we use this ratio to gain certain insights into the nature of the in-plane modes that will be relevant to analyzing their role in the broadening of the drumhead mode spectrum.

\subsection{\label{sec:therm_disp}Mean-squared thermal displacement}

In this Section, we will analyze the contribution of the two in-plane mode branches to the mean squared thermal displacement of the crystal, which is a measure of the fluctuations in the in-plane positions of the ions. From Eq.~(\ref{eqn:hperp}), the total energy in each mode decouples and therefore every complex amplitude $A_n$ can, in principle, be independently sampled from a Boltzmann distribution in order to study thermal properties at a temperature $T$. Accordingly, the mean energy in each mode is simply $k_B T$, independent of any details of the mode. In addition, we have the following thermal averages (denoted by an overbar) for the complex amplitudes $A_n$:
\begin{equation}
    \overline{A_n}=0,\; \overline{A_n A_k} = 0, \; \forall \; n,k
    \label{eqn:an_therm_1}
\end{equation}
and 
\begin{equation}
    \overline{A_n^* A_k} = \frac{k_B T}{\braket{u_n | \mathbb{E} | u_n}} \delta_{n,k}.
    \label{eqn:an_therm_2}
\end{equation}

In the crystal plane, the total squared displacement of all the ions is given by $\rho_\perp^2 = \braket{\delta r_\perp | \delta r_\perp}$. From Eq.~(\ref{eqn:q_harmonic}), we get 
\begin{equation}
    \overline{\rho_\perp^2} = 2\sum_{n=1}^{2N}  \frac{k_B T}{\braket{u_n | \mathbb{E} | u_n}} \braket{u_n^r | u_n^r},
\end{equation}
where we have used the thermal averages according to Eq.~(\ref{eqn:an_therm_1}) and Eq.~(\ref{eqn:an_therm_2}). Furthermore, the quantity $\braket{u_n | \mathbb{E} | u_n}$ can be expressed as 
\begin{equation}
    \braket{u_n | \mathbb{E} | u_n} = m\omega_n^2 \braket{u_n^r | u_n^r}\left(1 + R_n\right).
\end{equation}
We can now write 
\begin{equation}
    \overline{\rho_\perp^2} = \sum_{n=1}^{2N}  \overline{\rho_n^2},
\end{equation}
where $\overline{\rho_n^2}$ is the mean squared thermal displacement of mode $n$ and is given by 
\begin{equation}
        \overline{\rho_n^2} = \frac{2}{\left(1+R_n\right)} \frac{k_B T}{m\omega_n^2}.
\end{equation}
The simple factor $2/(1+R_n)$ relates the mean squared thermal displacement of an in-plane mode with that of a SHO ($R,2/(1+R)=1$) at the same frequency and temperature. 

In the $\exb$ branch, $R_n \sim \mathcal{O}(100)$ and therefore, $\overline{\rho_n^2}$ is typically $2$ orders of magnitude smaller than that of a SHO with the same $\omega_n$ and $T$. In the cyclotron branch, $R_n \approx 0$, and $\overline{\rho_n^2}$ is approximately twice that of a SHO with the same $\omega_n$ and $T$. 

We conservatively estimate the relative contribution of the $\exb$ and cyclotron branches to the total mean-squared thermal displacement of the crystal as follows. Selecting a representative mode for each branch, denoted respectively by the subscripts $m,c$, the ratio of $\overline{\rho_m^2}/\overline{\rho_c^2}$ is 
\begin{equation}
    \frac{\overline{\rho_m^2}}{\overline{\rho_c^2}} = \frac{1+R_c}{1+R_m} \frac{\omega_c^2}{\omega_m^2}.
\end{equation}
We can approximate $1+R_c\approx 1$ and $\omega_c/2\pi\approx 7.2$ MHz from Fig.~\ref{fig:n120_r_chi}(b). For a conservative estimate, from Fig.~\ref{fig:n120_r_chi}(a), we approximate $\omega_m/2\pi\approx 100$ kHz and $R_m \approx 200$, which leads to $\overline{\rho_m^2}/\overline{\rho_c^2} \approx 25$. Therefore, the mean squared displacement is predominantly due to the thermal energy in the $\exb$ modes. The $\exb$ branch contribution dominates despite the strong suppression arising from the factor of $2/(1+R_n)$. The reason is that $\overline{\rho_n^2}$ is inversely proportional to $\omega_n^2$, which is very small in this branch. 

\subsection{\label{sec:md_vk}Implications for molecular dynamics simulations}

In molecular dynamics simulations of the crystal dynamics, a common initialization method is to provide velocity kicks to the ions at a set temperature. Here we investigate the energy provided to each mode by such an initialization.

Assuming a temperature $T$, the in-plane initial velocities of the ions are sampled according to 
\begin{equation}
    \overline{v_{\perp,0}^{j,s}}=0; \; \overline{v_{\perp,0}^{j,s} v_{\perp,0}^{k,s'}} = \frac{k_B T}{m}\delta_{j,k}\delta_{s,s'},
    \label{eqn:vk_stat}
\end{equation}
where $s,s'=x,y$ and $j,k$ label the ions. Here, the overbar denotes an average over many realizations of the initial conditions. In the harmonic approximation, these velocities translate to initial complex amplitudes $A_n$ for the in-plane modes, which in a single realization are set according to
\begin{equation}
A_n = \frac{i m \omega_n \braket{u_n^r | v_{\perp,0} }}{\braket{u_n | \mathbb{E} | u_n}}.    
\end{equation}
Clearly, $\overline{A_n}=0$, while the average energy in mode $n$, given by $\mh_n = \overline{A_n^*A_n} \braket{u_n | \mathbb{E} | u_n}$, reduces to  
\begin{equation}
    \mh_n = \frac{k_B T}{1+R_n}.
\end{equation}
For a system of coupled simple harmonic oscillators, such as the crystal motion in the axial direction, $R_n=1$ for all the modes and such velocity kicks initialize each mode with an average energy of $k_B T/2$. Therefore, all the modes can be initialized at temperature $T$, i.e. with average energy $k_B T$, by simply providing velocity kicks at $2T$. Such an initialization can be used to ensure statistical properties for the position and velocity coordinates that are identical to the thermal values up to the second moments (see Appendix~\ref{app:vk_sho}).

For the in-plane modes, however, such velocity kicks are seen to initialize the two mode branches with very different energies. In the cyclotron branch, $R_n\approx 0$ and all the modes are initialized approximately with $k_B T$ as desired. However, in the $\exb$ branch, $R_n \sim 100$, and consequently the average energy in the modes is a factor $\sim 1/100$ smaller than $k_B T$.

In Sec.~\ref{sec:therm_disp}, we established that the dominant contribution to the mean squared thermal displacement of the crystal is from the thermal occupation of the $\exb$ modes. However, the velocity kick method hardly excites the $\exb$ modes and therefore results in an initial configuration with artificially suppressed ion position fluctuations. As we will demonstrate in Sec.~\ref{sec:therm_snap}, modeling the in-plane ion position fluctuations is critical to study the broadening effect of an in-plane temperature on the drumhead mode spectrum. Therefore, for our molecular dynamics simulations discussed in Sec.~\ref{sec:md_sim}, we supplement the velocity kick method with explicit initialization of the in-plane spatial coordinates via the Metropolis-Hastings (MH) algorithm \cite{pathria2011statistical}.  

For completeness, we end this section with a description of the initialization procedure we employ for our molecular dynamics simulations. Our starting point is  the zero temperature equilibrium crystal, wherein all the ions are at rest in the $z=0$ plane. We allow for different temperatures $T_\perp$ and $T_\parallel$, respectively, in the planar and axial directions. We can thereby study the variation of the drumhead spectrum at a fixed $T_\parallel$ while $T_\perp$ is varied.  

The velocities along the $x$ and $y$ directions are initialized using velocity kicks corresponding to temperature $T_\perp$. For the planar spatial coordinates, we employ the MH algorithm: From an existing crystal configuration, a new candidate configuration is generated by randomly displacing a single ion within a circle of constant radius centered on its current position. The new configuration is immediately accepted if its potential energy, given by $\Phi = \sum_{j=1}^N e\phi_j$ (see Eq.~(\ref{eqn:eff_pot_sim})), is lower than that of the old configuration. If the new potential energy is greater by a value $\Delta \Phi>0$, the new configuration is accepted with probability $e^{-\beta\Delta \Phi}$, where $\beta=1/(k_B T_\perp)$ is the usual Boltzmann factor. A single scan consists of repeating this procedure sequentially for every ion in the crystal. In turn, a large number of scans, $\gtrsim 1000$, are required to eventually achieve configurations that are typical of the set temperature. The initial configuration for this iterative procedure is the zero temperature equilibrium configuration. Compared to this configuration, typical configurations at a set temperature $T_\perp$ correspond to an average potential energy increase of approximately $k_B T_\perp$ per ion, accounting for the two spatial degrees of freedom. We enforce $z_j=0$ for all the ions while sampling the in-plane coordinates.

Finally, in the axial direction, the complex drumhead mode amplitudes are initialized with independent real and imaginary parts corresponding to temperature $T_\parallel$. The $z$ coordinates and velocities of the ions are then initialized as appropriate linear combinations of the mode amplitudes, with the linear transformation matrix constructed out of the mode eigenvectors. The drumhead mode frequencies and eigenvectors used in this procedure are computed by diagonalizing the stiffness matrix $\mathbb{K}_\parallel$, which in turn is obtained by linearizing the equations for $\dot{v}_j^z$ (see Eqs.~(\ref{eqn:mom_update}) and (\ref{eqn:linear_eom})) about the initial minimum energy configuration. 

\section{\label{sec:drumhead}Drumhead mode spectrum}

We are now ready to investigate the effect of in-plane thermal fluctuations on the drumhead mode spectrum. So far, we have worked  in the harmonic approximation, wherein the in-plane and out-of-plane motions decouple as can be seen from  Eq.~(\ref{eqn:linear_eom}). However, the Coulomb interaction is intrinsically anharmonic, and large amplitude in-plane motion eventually affects the spectrum of drumhead modes significantly. We will first present an intuitive picture for why it is reasonable to expect a poorly resolved drumhead mode spectrum as the in-plane temperature increases. Next, we will verify and improve our modeling of this effect using molecular dynamics simulations.  

\subsection{\label{sec:therm_snap}Intuitive thermal snapshot picture}

\begin{figure*}[!htb]
    \centering
    \includegraphics[width=0.8\textwidth]{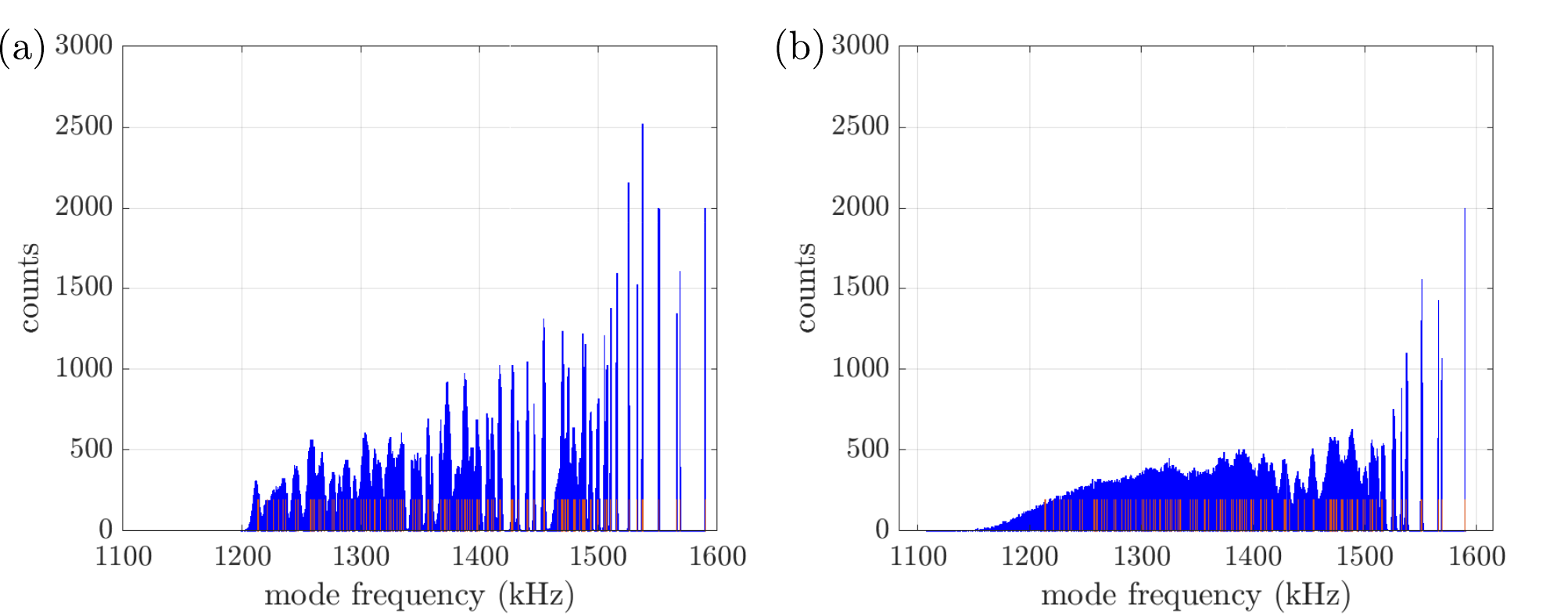}
    \caption{Histograms of sorted drumhead mode frequencies from $2000$ thermal snapshots of a crystal with $120$ ions, corresponding to an in-plane temperature of (a) $T_\perp = 1$ mK, and (b) $T_\perp = 10$ mK. The frequencies of the initial zero-temperature equilibrium configuration are overlaid for reference (orange). The histogram bin width is $500$ Hz. Relevant trap and crystal parameters are reported in the caption of Table~\ref{tab:char_freq}.}
    \label{fig:N120_histogram}
\end{figure*}

\begin{figure*}[!htb]
    \centering
    \includegraphics[width=0.8\textwidth]{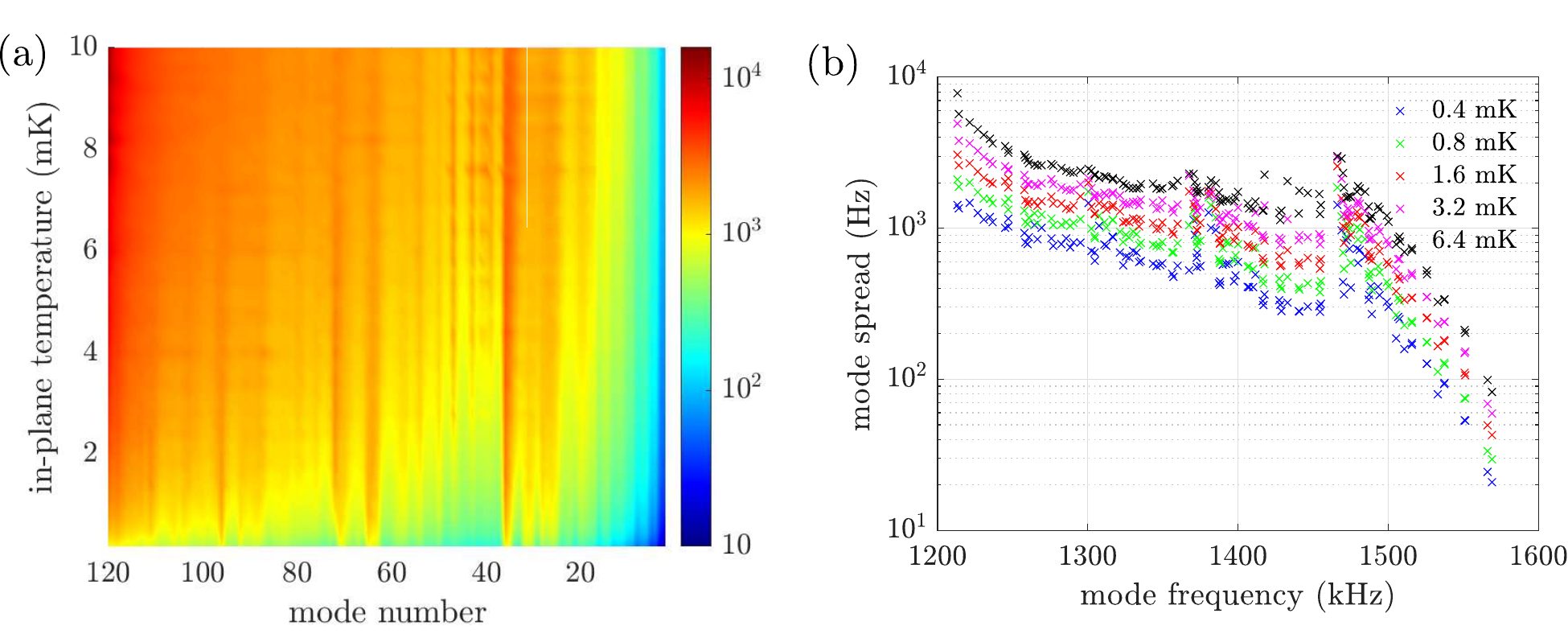}
    \caption{Investigation of mode frequency fluctuations. (a) Surface plot showing the standard deviation (in Hz) of frequency fluctuations as a function of mode number as the in-plane temperature is varied. For each temperature, 2000 thermal snapshots were generated to aggregate statistics. (b) Horizontal cuts at constant values of $T_\perp$ showing the standard deviation of frequency fluctuations, now as a function of mode frequency. We have used the sorted mode frequencies of the initial zero temperature equilibrium configuration for translating mode numbers into frequency values on the $x$ axis. In both panels, the c.m. mode (mode number $n=1$) has been left out because, for a single-species crystal, its frequency is insensitive to in-plane thermal fluctuations and is fixed at the axial trapping frequency $\omega_\parallel$. Relevant trap and crystal parameters are reported in the caption of Table~\ref{tab:char_freq}.}
    \label{fig:N120_stdval}
\end{figure*}

In-plane ion position fluctuations are essentially due to the low frequency $\exb$ motion, as seen by the contribution of this branch of modes to the mean-squared thermal displacement of the crystal. Therefore, for the present analysis, we ignore the fast cyclotron motion since their contribution to position fluctuations is not significant. During a single time period associated with a typical $\exb$ mode frequency ($\sim 10 - 100$ kHz), every drumhead mode ($\gtrsim 1$ MHz) completes several oscillations. In an extreme simplification, we imagine the ions to be thermally displaced but stationary in plane, while they execute rapid out-of-plane motion. We refer to such a stationary in-plane configuration as a thermal snapshot. We then follow the procedure of Ref.~\cite{wang2013PRA} to perform a normal mode analysis in the out-of-plane direction and sort the drumhead modes in descending order. We then repeat this exercise for a large number of thermal snapshots that are typical of the set in-plane temperature and histogram the sorted drumhead modes into frequency bins to study the spread in mode frequencies.

Thermal snapshots are generated by Metropolis-Hastings sampling of the in-plane spatial coordinates, as discussed in Sec.~\ref{sec:md_vk}. For a fixed value of  $T_\perp$, we perform a normal mode analysis on 2000 thermal snapshots. Successive snapshots are chosen $100$ scans apart to ensure that they are sufficiently uncorrelated. Figure~\ref{fig:N120_histogram} shows the histograms of drumhead frequencies for $T_\perp = 1$ mK and $10$ mK. For reference, the normal mode spectrum of the initial zero temperature equilibrium configuration is overlaid on these histograms. As $T_\perp$ increases, the spectrum transitions into a smooth and continuous distribution where only the first few modes are well-resolved. These histograms reflect the broadening effect of in-plane ion position fluctuations on the drumhead spectrum. 

To better understand the features of these histograms, we investigate the spread in mode frequency as a function of $T_\perp$ and the mode number $n$. For each thermal snapshot, we label the modes such that the index $n$ increases with decreasing mode frequency. Figure~\ref{fig:N120_stdval}(a) shows the standard deviation in mode frequencies as a function of mode number, as the in-plane temperature $T_\perp$ is varied from $200 \; \mu$K to $10$ mK in steps of $200 \; \mu$K. We observe that the highest frequency modes are fairly robust to $T_\perp$ and only experience frequency fluctuations in the range of a few $100$ Hz even around $T_\perp = 10$ mK. In contrast, the lowest frequency modes are highly sensitive to $T_\perp$ and experience frequency fluctuations in the range of $10$ kHz as $T_\perp$ increases. Although the general tendency is an increase in frequency fluctuations at higher mode numbers (lower mode frequencies), a remarkable feature is the appearance of bands  of modes at intermediate frequencies that are significantly more sensitive to $T_\perp$ than nearby modes. These features can also be inferred from Fig.~\ref{fig:N120_stdval}(b), where we plot horizontal cuts of the surface plot shown in Fig.~\ref{fig:N120_stdval}(a) for fixed $T_\perp$ values. For a number of modes, the standard deviation in their frequencies is proportional to $\sqrt{T_\perp}$. This trend is particularly easy to observe in the case of the well-resolved high frequency modes. 

The variation in the sensitivity of the drumhead modes to in-plane thermal fluctuations can be partially explained by the following intuitive argument. A highly delocalized mode, i.e. one that is supported by a large number of ions, can be expected to be fairly insensitive to a random fluctuation in the position of any single ion. In contrast, a highly localized mode that is supported only by a small number of ions is likely to be very sensitive to a fluctuation in the position of one of these ions. To quantify the extent of delocalization of a drumhead mode, we introduce the mode entropy $H_n^\parallel$. In terms of the orthonormal drumhead mode eigenvectors, which are denoted by $\ket{b_n^\parallel}$ and whose elements $b_{jn}^\parallel$ quantify the displacement of ion $j$ as a result of mode $n$, the entropy for mode $n$ is defined as 
\begin{equation}
    H_n^\parallel = -\sum_{j=1}^N \abs{b_{jn}^\parallel}^2 \log_{2} \abs{b_{jn}^\parallel}^2.
\end{equation}
This definition is inspired from that of the Shannon entropy in information theory. Such a definition is enabled by the realization that, since $\sum_{j=1}^N \abs{b_{jn}^\parallel}^2 = 1$, the squared mode amplitudes can be interpreted as probabilities for selecting an ion $j$ based on its contribution to the total power in mode $n$. For a highly delocalized mode, such as the c.m. mode, $\abs{b_{j,1}^\parallel}^2=1/N\; \forall \; j=1,\ldots,N$ and $H_1^\parallel = \log_{2} N$. For a hypothetical mode $n'$ that is only supported by one ion $j'$, $H_{n'}^\parallel = 0$, where we have evaluated the entropy in the limit that $\abs{b_{j',n'}^\parallel}^2\rightarrow 1$ while $\abs{b_{j,n'}^\parallel}^2\rightarrow 0 \; \forall \; j \neq j'$.  

\begin{figure}[!tb]
    \centering
    \includegraphics[width=0.9\columnwidth]{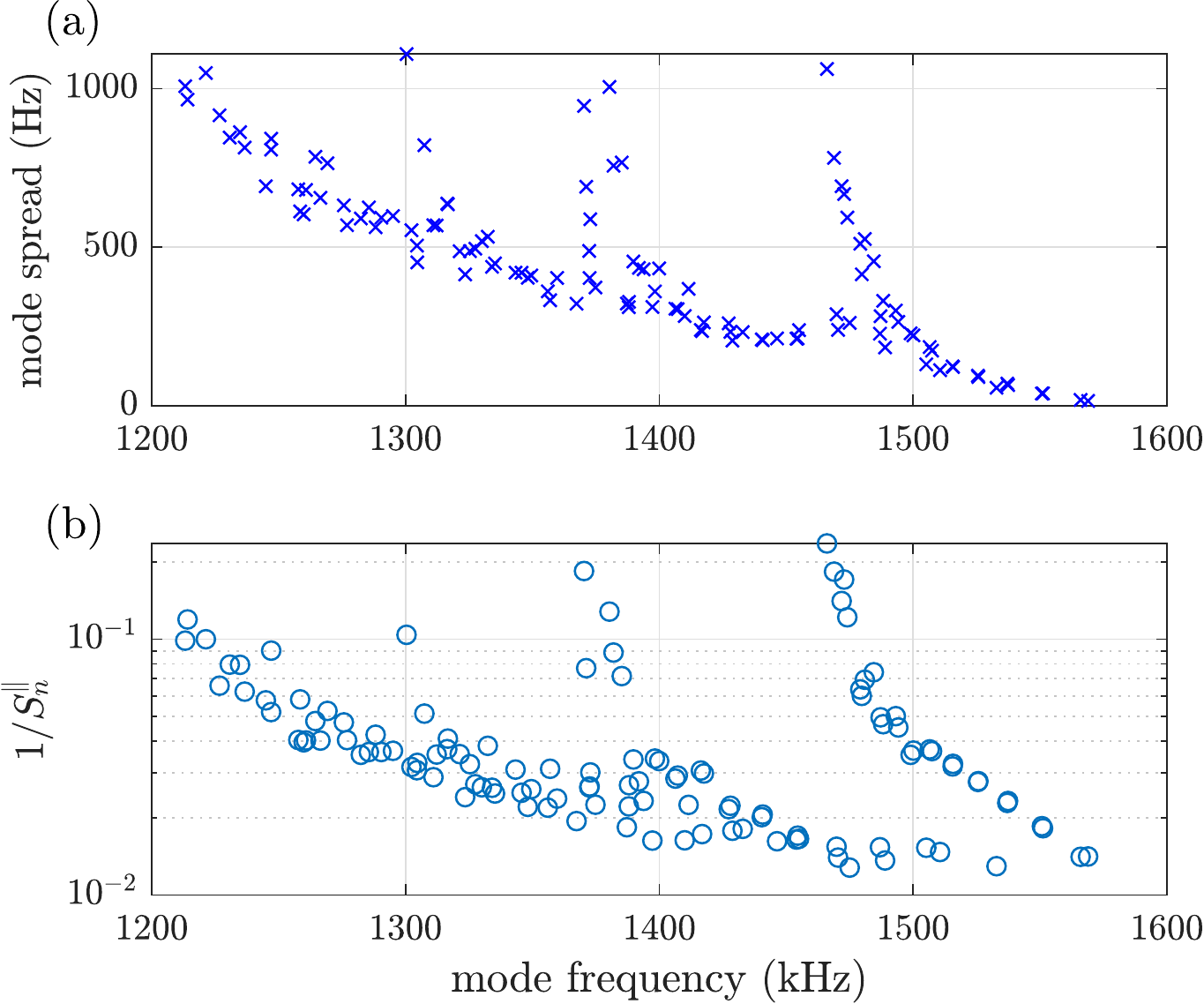}
    \caption{Correlation between frequency fluctuations and support number. (a) Standard deviation of frequency fluctuations as a function of mode frequency at $T_\perp = 200 \; \mu$K. (b) Inverse support number as a function of mode frequency (see Eq.~(\ref{eqn:sp_num})). Once again, the c.m. mode is omitted. Relevant trap and crystal parameters are reported in the caption of Table~\ref{tab:char_freq}.} 
    \label{fig:N120_mode_metrics}
\end{figure}

For better physical intuition, we can convert this entropy into an effective number of ions supporting a mode. We define the support number $S_n^\parallel$ for mode $n$ as 
\begin{equation}
    S_n^\parallel = 2^{H_n^\parallel}.
    \label{eqn:sp_num}
\end{equation}
The support number is independent of the choice of base, chosen here to be $2$, as long as the same base is used in calculating $H_n^\parallel$. The c.m. mode has the maximum support number given by $S_1^\parallel = N$, implying that it is supported by all the ions in the crystal, and hence, it is the most delocalized mode. For our hypothetical mode supported by a single ion, we recover $S_{n'}^\parallel = 1$ as expected. All the other drumhead modes have support numbers in between these two extremes. 

In Fig.~\ref{fig:N120_mode_metrics}, we investigate the correlation between frequency fluctuations and the support number. Figure~\ref{fig:N120_mode_metrics}(a) shows the standard deviation of frequency fluctuations as a function of mode frequency at $T_\perp = 200 \; \mu$K. In Fig.~\ref{fig:N120_mode_metrics}(b), we plot the inverse of the support number versus the mode frequency. The mode amplitudes used for this calculation correspond to the drumhead modes of the zero temperature equilibrium configuration. The striking similarity between the two panels validates our hypothesis that a poorly supported mode is generally more sensitive to in-plane thermal fluctuations. Figure~\ref{fig:N120_mode_metrics}(b) reveals the existence of bands of poorly supported modes at intermediate frequencies. In turn, these bands are more sensitive to $T_\perp$ and exhibit higher levels of frequency fluctuations compared to neighboring modes that are better supported.

\subsection{\label{sec:md_sim}Molecular dynamics simulations}

We now proceed to investigate the effect of in-plane thermal fluctuations on the drumhead mode spectrum using molecular dynamics simulations. We first initialize the position and velocity degrees of freedom according to the discussion in Section~\ref{sec:md_vk}. Next, we use a 4th order Runge-Kutta integrator to numerically evolve the equations of motion given by Eq.~(\ref{eqn:pos_update}) and Eq.~(\ref{eqn:mom_update}). Observables are averaged over $96$ realizations of the crystal dynamics, in order to account for the random nature of the thermal initial conditions. The total simulation time in each realization is $\mathcal{T}_\text{tot} = 560 \; \mu$s, which is a typical time taken for a single experimental sequence probing the drumhead spectrum (see Section~\ref{sec:odf}). The crystal energy is expected to be constant over time and fluctuations in this quantity can be considered as a measure of the numerical accuracy. We have found that the energy fluctuates at a fractional level $<2\times10^{-6}$ relative to the total crystal energy. Alternatively, in terms of the thermal energy of the crystal, which is the excess over the potential energy of the equilibrium configuration, the fractional energy fluctuations do not exceed $2\times 10^{-3}$.

\subsubsection{Power spectral density}

\begin{figure*}[!tb]
    \centering
    \includegraphics[width=\textwidth]{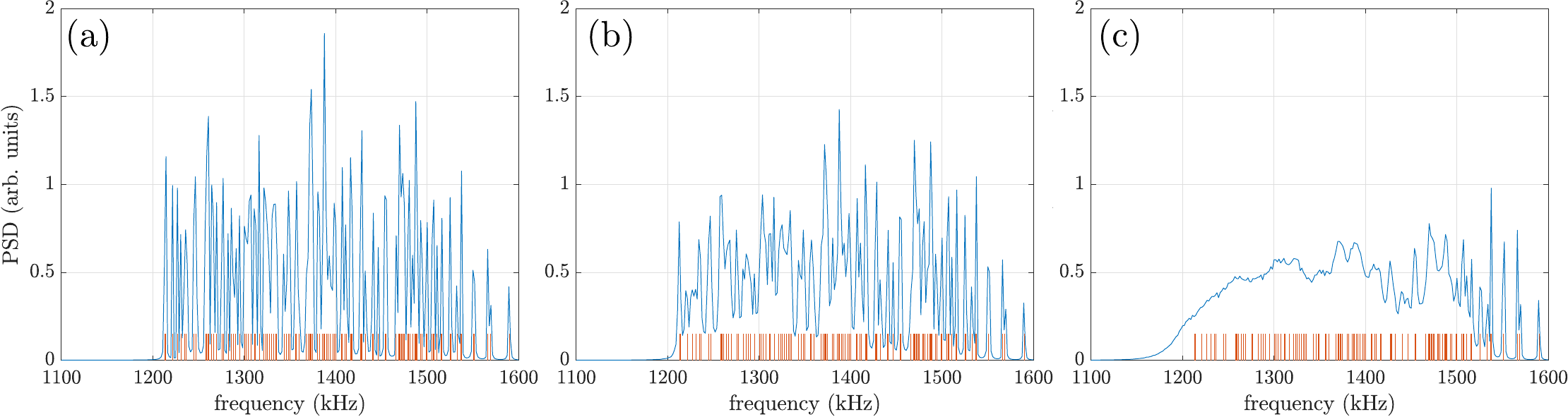}
    \caption{Power spectral density (PSD) of the drumhead motion for a $120$ ion crystal with (a) $T_\perp = 0$, (b) $T_\perp = 1$ mK and (c) $T_\perp = 10$ mK. The frequencies predicted from a normal mode analysis of the zero-temperature equilibrium configuration are overlaid for reference (orange). Relevant trap and crystal parameters are reported in the caption of Table~\ref{tab:char_freq}. }
    \label{fig:N120_ps}
\end{figure*}

\begin{figure*}[!htb]
    \centering
    \includegraphics[width=0.8\textwidth]{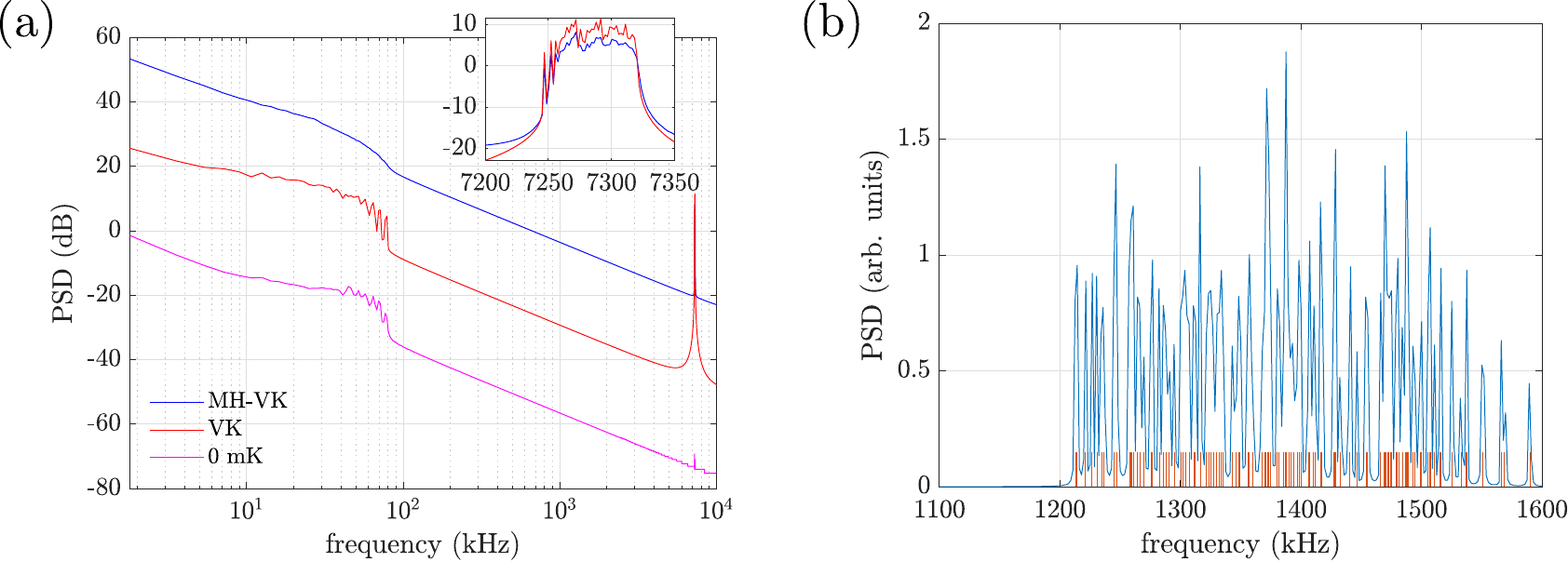}
    \caption{Importance of Metropolis-Hastings sampling of in-plane spatial coordinates. (a) The red curve shows the PSD (in dB) of the in-plane motion when it is initialized using only velocity kicks (VK) at twice the in-plane temperature of $T_\perp=10$ mK, while the blue curve is the PSD when the motion is initialized using a combination of velocity kicks and Metropolis-Hastings sampling (MH-VK) of the planar spatial coordinates (both implemented with temperature $T_\perp$). Inset: Magnified view of the PSD in the VK and MH-VK cases in the cyclotron branch. The main panel also shows the in-plane PSD when $T_\perp = 0$ mK for reference. This PSD is not identically zero possibly because of (i) the finite accuracy in obtaining the zero temperature equilibrium configuration and in the subsequent numerical time evolution, (ii) the MH sampling, which may occasionally find lower-energy configurations than the initial input configuration, and (iii) nonlinear coupling between the out-of-plane and in-plane degrees of freedom. (b) PSD of the drumhead motion in the VK case.  Relevant trap and crystal parameters are reported in the caption of Table~\ref{tab:char_freq}.  }
    \label{fig:N120_vk}
\end{figure*}

We analyze the power spectral density (PSD) of the out-of-plane and in-plane motion to investigate the broadening effect and its physical origin. The details of the PSD computation are presented in Appendix~\ref{app:psd}.

Figure~\ref{fig:N120_ps} shows the power spectral density (PSD) of the drumhead motion in a $120$ ion crystal for three different in-plane temperatures, $T_\perp = 0, 1$ mK  and $10$ mK. The out-of-plane temperature is assumed to be $T_\parallel = 0.5$ mK, which results in a mean thermal occupation of $\nbar_{\text{c.m.}} \approx k_B T_\parallel/(\hbar \omega_\parallel) \approx 6.5$ phonons in the c.m. mode. This value is slightly larger than typical c.m. mode occupations measured after Doppler cooling \cite{jordan2019PRL}. 

Figure~\ref{fig:N120_ps}(a) shows that the PSD is fairly well-resolved in the absence of in-plane motion. The location of the peaks are consistent with the frequencies expected from a normal mode analysis (orange). As $T_\perp$ is increased to $1$ mK, sharp features begin to broaden out as depicted in Fig.~\ref{fig:N120_ps}(b). Further increasing $T_\perp$ to $10$ mK (panel (c)), we observe that the PSD becomes a smooth continuum over modes and only the first few highest frequency modes are well-resolved. The qualitative features observed in the $10$ mK spectrum are consistent with experimental measurements of the drumhead spectrum.

To verify the importance of Metropolis-Hastings sampling of the in-plane spatial coordinates, we now assume that the in-plane degrees of freedom can be initialized analogous to simple harmonic oscillators, i.e. by providing velocity kicks at twice the in-plane temperature, chosen here to be $T_\perp=10$ mK (see Section~\ref{sec:md_vk}). In Fig.~\ref{fig:N120_vk}(a), we compare the PSD (in dB) of the in-plane motion resulting from an initialization employing only velocity kicks at $2T_\perp$ (red, henceforth referred to as VK) versus the PSD ensuing from the combined initialization using velocity kicks and Metropolis-Hastings sampling of the in-plane spatial coordinates, both of which are implemented with temperature $T_\perp$ (blue, referred to as MH-VK). First, the difference between the two PSD profiles is at least $15$ dB over the entire frequency range from $0$ to $100$ kHz. Therefore, the $\exb$ branch in the VK case carries at least $\sim 30$ times lower power than in the MH-VK case. Second, as seen in the inset, the VK PSD is roughly $3$ dB higher than the MH-VK PSD in the cyclotron branch, corresponding to twice as much power in this frequency range. Both of these observations are consistent with our conclusion in Section~\ref{sec:md_vk} that velocity kicks predominantly initialize only the cyclotron branch and poorly initialize the $\exb$ branch; since such kicks are provided at $2T_\perp$ in the VK case, the PSD is correspondingly larger by a factor of $2$ in the cyclotron branch.     

Figure~\ref{fig:N120_vk}(b) shows the drumhead PSD at $T_\perp = 10$ mK in the VK case. In contrast to Fig.~\ref{fig:N120_ps}(c), the spectrum is now well resolved and in fact resembles the $T_\perp = 0$ mK case (Fig.~\ref{fig:N120_ps}(a)), confirming that the broadening of the drumhead spectrum can be attributed to energy in the $\exb$ branch. Combined with the result that modes in this low-frequency branch are excited almost exclusively by MH sampling of the spatial coordinates, this observation also lends further support to our intuitive thermal snapshot picture discussed in Section~\ref{sec:therm_snap}. 

\subsubsection{\label{sec:odf}ODF spectrum}

Experimental observations of the drumhead mode spectrum are enabled by the application of an optical dipole force (ODF). The experimental setup to realize the ODF and its physical interpretation in the context of thermometry have been described in detail elsewhere, for example, see Refs.~\cite{sawyer2012PRA,sawyer2012PRL,shankar2020Thesis}. For our present purposes, we recall that this interaction engineers a motion-induced dephasing of an equal superposition of two hyperfine levels constituting a pseudospin-1/2 system. The ODF interaction Hamiltonian for ion $j$ is 
\begin{equation}
    \hat{H}_\text{ODF}^j = F_0 \cos\left(\mu_R t\right) \hat{z}_j \hSig_j^z,
\end{equation}
where $F_0$ is the ODF magnitude, $\mu_R$ is the difference frequency of the two lasers used to generate this interaction, $\hat{z}_j$ is the $z$-position operator for ion $j$, and $\hSig_j^z$ is the Pauli matrix for the pseudospin-1/2 system. 

\begin{figure}[!tb]
    \centering
    \includegraphics[width=\columnwidth]{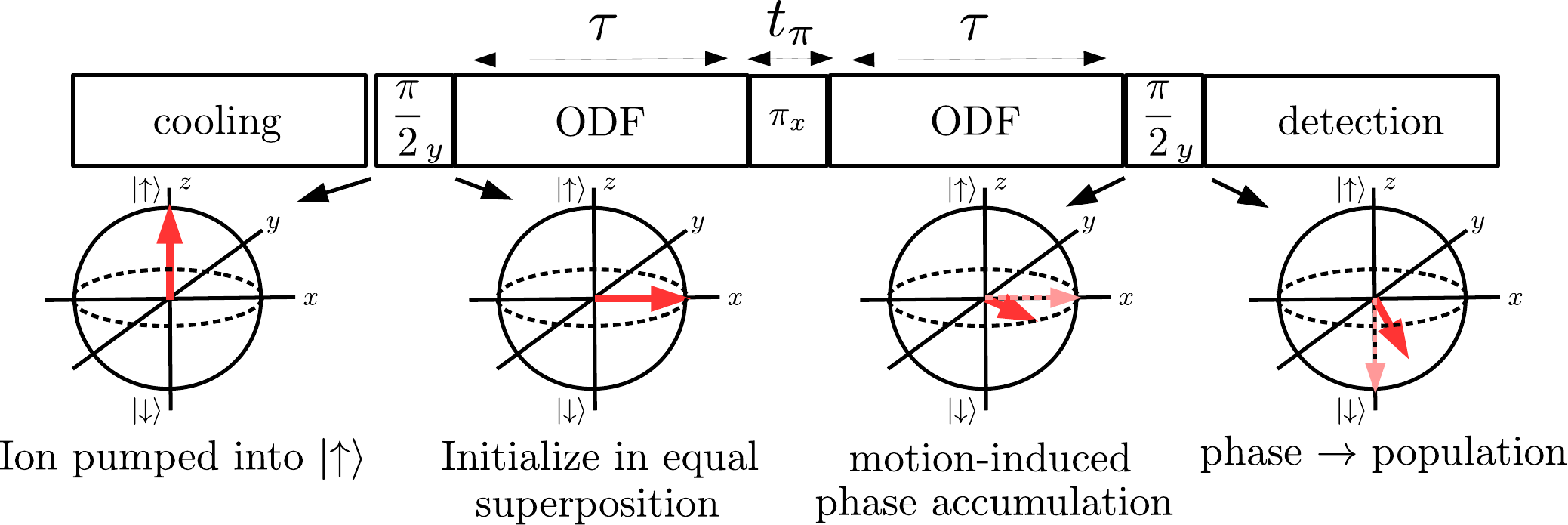}
    \caption{Experimental sequence for thermometry using the optical dipole force (ODF). Bloch spheres qualitatively depict the pseudospin state of any single ion at various stages of the thermometry protocol.}
    \label{fig:odf_bloch}
\end{figure}

Fig.~\ref{fig:odf_bloch} depicts the thermometry sequence along with Bloch sphere visualizations of any single pseudospin at various stages of this protocol. The protocol is a Ramsey-like sequence where the pseudospin is first initialized in an equal superposition of the two spin states using a $\pi/2$ pulse. Next, the ODF couples the motion and spin degrees of freedom for a total time $2\tau$ with a spin-echo pulse of duration $t_\pi$ sandwiched in between. The echo cancels additional dephasing such as that arising from slowly varying magnetic field inhomogeneities. Under the influence of the ODF, the Bloch vector accumulates a phase that is dependent on the ion motion. The contribution of specific motional frequency components to the accumulated phase can be near resonantly enhanced by adjusting the ODF difference frequency $\mu_R$ close to these components. The final $\pi/2$ pulse maps the azimuthal angle on to the polar angle in the Bloch sphere, and therefore, the phase accumulated can be measured via the population in the $\ket{\uparrow}$ spin state, which is the bright state. For thermal motion, the phase accumulated in each repetition of the sequence varies, resulting in effective motion-induced spin dephasing \cite{sawyer2012PRA,shankar2020Thesis} when averaged over all the repetitions. A plot of the fraction of ions in the bright state versus the ODF difference frequency $\mu_R$ as the latter is varied over the full bandwidth of drumhead modes is called the ODF spectrum. Such a plot gives qualitative information about the energy content at various frequencies and can be used to quantitatively estimate the temperature of well resolved drumhead modes.   

\begin{figure*}[!tb]
    \centering
    \includegraphics[width=\textwidth]{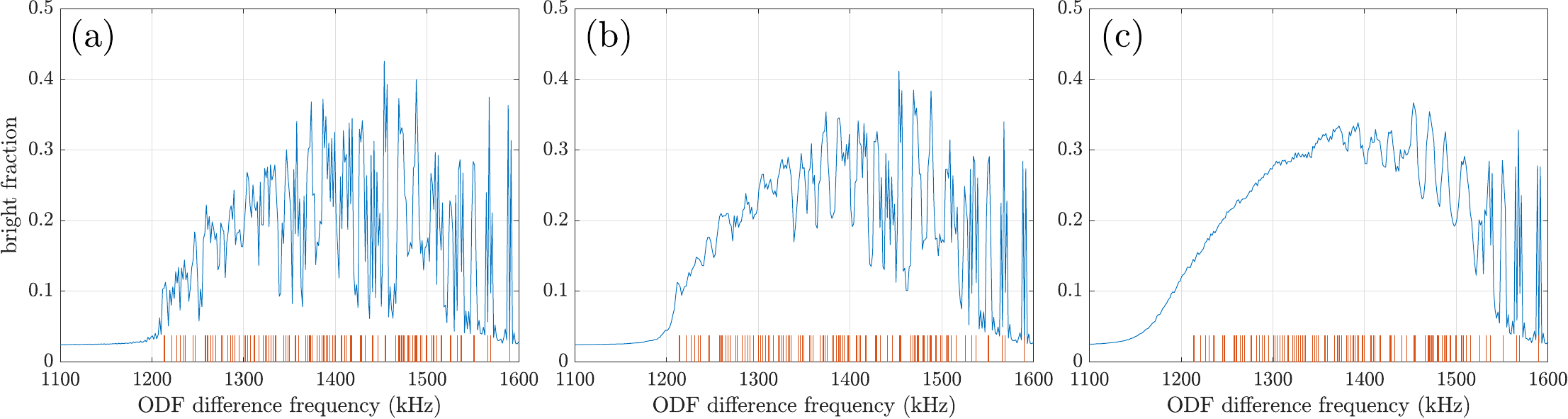}
    \caption{Simulated ODF spectrum corresponding to the drumhead motion of a $120$-ion crystal with (a) $T_\perp = 0$, (b) $T_\perp = 1$ mK and (c) $T_\perp = 10$ mK. The frequencies predicted from a normal mode analysis are overlaid for reference (orange). Relevant trap and crystal parameters are reported in the caption of Table~\ref{tab:char_freq}.}
    \label{fig:N120_odf_spec}
\end{figure*}

The objective in this Section is to demonstrate how the ODF spectrum can be estimated from molecular dynamics simulations and then study qualitative features of the simulated ODF spectra. To do so, we investigate the effect of $\hat{H}_\text{ODF}^j$ on the motion and spin dynamics in a mean field approximation. In addition to the trap dynamics (Eq.~(\ref{eqn:mom_update})), the equation of motion for $v_j^z$ picks up an additional acceleration $a_j^{\text{ODF}}$ given by
\begin{equation}
     a_j^{\text{ODF}} = -\frac{F_0}{m} \cos\left(\mu_R t\right) \ev{\hSig_j^z}.
\end{equation}
However, $\ev{\hSig_j^z} \approx 0$ since all the ions are initialized in an equal superposition by the first $\pi/2$ pulse. Therefore, the effect of the ODF on the ion motion vanishes in the mean field approximation. 

For the quantity $\ev{\hSig_j^+}=\ev{\hSig_j^x}/2+i\ev{\hSig_j^y}/2$, the expectation value of the Pauli excitation operator, we get 
\begin{equation}
    \frac{d}{dt}\ev{\hSig_j^+} = \frac{2iF_0}{\hbar} \cos\left(\mu_R t \right) z_j \ev{\hSig_j^+},
\end{equation}
where we have replaced $\ev{\hat{z}_j}$ by the classical coordinate $z_j$. Since the first $\pi/2$ pulse prepares ions along the $x$ axis of the Bloch sphere, $\ev{\hSig_j^+(0)}=1/2$, and its time evolution is then given by $\ev{\hSig_j^+(t)} =  \exp \left(i A_j(t) \right)/2$, where
\begin{equation}
    A_j(t) = \frac{2F_0}{\hbar}\int_0^t dt' z_j(t') \cos\left(\mu_R t'\right).
    \label{eqn:acc_phase}
\end{equation}
The effect of the spin echo pulse can be accounted for by introducing an additional modulation $g(t)$ in the integrand appearing in Eq.~(\ref{eqn:acc_phase}). The function $g(t)$ takes the value $+1$ when $0<t<\tau$, $0$ when $\tau<t<\tau+t_\pi$ (ODF lasers switched off), and $-1$ when $\tau+t_\pi<t< \mathcal{T}_\text{tot}\equiv2\tau+t_\pi$ (to account for the sign change in the $y$ component of the spin after the echo pulse).

The final $\pi/2$ pulse maps the azimuthal angle on to the polar angle, i.e. $-\ev{\hSig_j^x}\rightarrow\ev{\hSig_j^z}$. After this pulse, the probability of finding an ion $j$ in the bright state, given by $(1+\ev{\hSig_j^z})/2$, is 
\begin{equation}
    P_j\left(\uparrow\right) = \frac{1}{2}\left(1 - e^{-2\Gamma\tau}\cos \left(A_j(2\tau+t_\pi)\right) \right).
    \label{eqn:pbright_md}
\end{equation}
In writing Eq.~(\ref{eqn:pbright_md}), we have accounted for additional dephasing arising from off-resonant scattering of the ODF lasers at rate $\Gamma$  by the ions \cite{sawyer2012PRL,jordan2019PRL}. Rather than the population of any single ion, the experimental observable is the total fraction of ions in the bright state,  measured by detecting the net fluorescence intensity from the crystal.

Because of the mean-field approximation and the classical treatment of ion motion, the ODF spectrum estimated in this manner does not capture the dephasing arising from zero-point motion and the phonon-mediated spin-spin interactions \cite{shankar2020Thesis}. However, for a first qualitative analysis, both of these effects may be assumed as small corrections when the mean phonon occupations in the drumhead modes are large, which is indeed the case after Doppler cooling.

In Fig.~\ref{fig:N120_odf_spec}, we qualitatively compare the simulated ODF spectrum for three different values of the in-plane temperature, namely, $T_\perp = 0, 1$ mK and $10$ mK. Similar to the PSD, we find that the ODF spectrum partially resolves the drumhead modes in the absence of in-plane thermal fluctuations but broadens into a poorly resolved spectrum as $T_\perp$ increases. Experimentally measured ODF spectra are qualitatively most similar in appearance to the simulated spectrum at $T_\perp = 10$ mK (see Fig.~\ref{fig:n53_odf_paper}(b)). Together with our conclusion that thermal energy in the $\exb$ modes are responsible for drumhead frequency fluctuations, this observation indicates that Doppler cooling of the $\exb$ motion is not very efficient under the current operating conditions of the NIST Penning trap.

\section{\label{sec:conc}Conclusion and outlook}

We have demonstrated that in-plane thermal fluctuations of ion positions have a strong broadening effect on the spectrum of drumhead modes in two-dimensional crystals stored in Penning traps.  These position fluctuations can be predominantly attributed to thermal energy in the low frequency $\exb$ modes. The energy in these modes is predominantly potential energy arising by virtue of the ion displacements from equilibrium, whereas the energy in the high frequency cyclotron modes is predominantly kinetic and arises by virtue of ion motion. We demonstrated this unconventional feature, arising from the presence of a strong magnetic field, by showing that the ratio $R_n$ of time-averaged potential to kinetic energy in a mode is $\gg 1$ ($\ll 1$) for the $\exb$ (cyclotron) modes. For current parameters of the NIST Penning trap, these ratios are typically in the range $R_n \sim 200$ ($R_n \sim 0.005)$ for the $\exb$ (cyclotron) branches. From  the perspective of molecular dynamics simulations, these values imply that initializing the crystal by providing velocity kicks to the ions largely provides energy input only to the cyclotron modes, whereas the $\exb$ branch remains largely unexcited. To overcome this problem, we proposed to supplement velocity kicks by explicit initialization of the in-plane spatial coordinates using the Metropolis-Hastings algorithm. 

To understand the effect of these fluctuations on the drumhead modes, we first studied the histogram of drumhead mode frequencies obtained from a normal mode analysis of an ensemble of static thermal snapshots of the in-plane ion crystal configuration. We observed that the histogram broadens as fluctuations in the ion positions increase with increasing in-plane temperature. We qualitatively explained the sensitivity of the frequency of any single drumhead mode to in-plane fluctuations by introducing a metric called the support number. This metric quantifies the extent to which a mode is delocalized over a crystal and generally revealed that the frequency of a highly delocalized (localized) mode is less (more) sensitive to in-plane fluctuations. 

Subsequent to this intuitive analysis, we verified the broadening of the drumhead spectrum in the presence of in-plane thermal fluctuations by carrying out molecular dynamics simulations of the crystal dynamics. We verified that the resolution of the drumhead spectrum degrades with increasing in-plane temperature. We also demonstrated that the broadening effect cannot be observed if the in-plane degrees of freedom are initialized using only velocity kicks and that additional MH initializiation of spatial coordinates is required. By studying the in-plane power spectrum, we showed that the velocity kicks predominantly provide energy to the cyclotron branch and leave the $\exb$ branch poorly initialized. 

While the drumhead power spectrum is a useful simulation diagnostic, it does not directly correspond to an observable in a real experiment. Rather, the drumhead spectrum is probed by mapping the ion motion on to the population in an electronic level using the optical dipole force (ODF). Therefore, we demonstrated how the ODF spectrum can be extracted from molecular dynamics simulations for direct comparison to experimentally measured spectra. By observing the simulated ODF spectra at various in-plane temperatures, we concluded that in-plane temperatures of the order of $T_\perp \sim 10$ mK are required to obtain spectra as poorly resolved as that observed in experiments. In the future, the ODF spectrum diagnostic can be incorporated into simulations that explicitly model the Doppler cooling process \cite{tang2019PoP} so that the simulated spectra can be directly compared to the experimental data. Through such comparisons, we are interested to see if the broadening of the drumhead spectrum can be used to indirectly estimate the in-plane temperature, which is an observable that has so far proved difficult to measure.

Previous studies have investigated the efficiency of perpendicular Doppler laser cooling on in-plane motion and have indicated that in-plane temperatures as low as $T_\perp \approx 1$ mK may be possible \cite{torrisi2016PRA,tang2019PoP}. However, we emphasize that these studies define the temperature based on only the kinetic energy of the in-plane motion and do not consider thermal potential energy fluctuations. Therefore, their conclusions only concern the temperature of the cyclotron modes. On the other hand, Doppler cooling may not be very efficient in removing potential energy from the $\exb$ modes that likely carry energy of the order of $10$ mK per mode. 

Our study therefore motivates the need for techniques that improve the cooling of the $\exb$ modes. For single ions in Penning traps, improved cooling of the $\exb$ motion has been proposed and demonstrated by applying a resonant drive that couples this mode to either the cyclotron mode or the out-of-plane mode, both of which are efficiently Doppler cooled. Recently, such `axialization' techniques have been used to cool the $\exb$ motion of a single ion down to the regime where sideband cooling can subsequently be used to further cool this mode to its ground state \cite{hrmo2019PRA}.\footnote{In these studies pertaining to single trapped ions, cooling is defined as a reduction in motion amplitude, which for the single $\exb$ or magnetron mode corresponded to an increase of mode energy. An important distinction of our system is the presence of a rotating wall. Throughout our work, we have assumed that the wall strength is strong enough to ensure that the rigid rotation of the crystal is locked to the wall frequency $\omega_r$. Therefore, the natural frame to analyze the normal modes of the crystal is the frame co-rotating with the ion crystal. In this frame, the $\exb$ modes have the more conventional property that a reduction in mode amplitude is associated with a reduction in mode energy.} Axialization has also recently been theoretically demonstrated to improve the cooling of $\exb$ modes in small ion crystals stored in a Penning trap array \cite{jain2018arXiv}. Future investigations of axialization techniques may also open avenues for improved cooling of $\exb$ modes in large crystals stored in a single Penning trap. A related possibility is to simply increase the rotating wall frequency so that the bandwidth of the drumhead modes increases sufficiently to partly overlap with the $\exb$ modes. Even in the absence of an additional drive, the Coulomb interaction will then cause energy exchange with the drumhead modes and potentially lead to more efficient cooling of the $\exb$ modes. 

A different attempt at directly improving the efficiency of Doppler cooling involves operating in a regime where the in-plane modes are less strongly magnetized. Such a regime can be accessed by lowering the cyclotron frequency, either by decreasing the magnetic field or by using a heavier mass ion. The motivation is that the value of $R_n$ for the $\exb$ modes will then be closer to unity, and cooling of these modes will then be more similar to the Doppler cooling of simple harmonic oscillator modes. Another more speculative idea is the use of a co-rotating `tweezer' beam (or beams) near the crystal boundary for stabilizing the crystal structure and possibly mitigating the impact of potential energy fluctuations \cite{laupretre2020PRA}.

Other processes, such as rearrangements of the ion crystal from background gas collisions, could also possibly contribute to the broadening of the drumhead mode spectrum. Nevertheless, our study indicates that a non-zero in-plane temperature is an important source of drumhead spectral broadening and emphasizes the need for improved cooling of the $\exb$ modes, which will undoubtedly have more general benefits.  For instance, reducing the in-plane position fluctuations greatly improves the prospects for site-resolved addressing of ions in these two-dimensional crystals, further enhancing the range of quantum information protocols that can be implemented on such devices.

\acknowledgments{We thank Francois Anderegg, Bjorn Sumner and Brian Sawyer for useful comments and discussions on the manuscript. A.S. thanks Akshay Seshadri for help with the python coding language. This work was partially supported by NSF PFC Grant No. PHY 1734006, the DARPA and ARO Grant No. W911NF-16-1-0576, DARPA ONISQ, the AFOSR Grant No. FA 99550-20-1-0019, a DOE Office of Science HEP QuantISED award, and a U.S. Department of Energy grant DE-SC0020393.  M.A. was supported by an NRC fellowship funded by NIST. D.D. acknowledges support from AFOSR contract FA 9550-19-1-0999, DOE Grant No. DE-SC0018236, and NSF Grant No. PHY1805764.}

\appendix

\section{\label{app:vk_sho}Velocity kick initialization of a simple harmonic oscillator}

Frequently velocity kicks are used to realize a thermal ensemble.  Here we show that velocity kicks followed by a random mixing time does not strictly realize a thermal ensemble.

We consider a simple harmonic oscillator described by the Hamiltonian 
\begin{equation}
    \mh = \frac{1}{2}m v^2 + \frac{1}{2}k x^2,
\end{equation}
written in terms of the position $x$ and velocity $v$. We consider two methods to initialize the oscillator at temperature $T$ and compare the resulting moments.

\subsection{Method 1}

We independently sample the position and velocity according to Gaussian distributions with $\overline{x}=0, \overline{v}=0$ and variances 
\begin{equation}
    \overline{x^2} = \frac{k_B T}{k},\; 
    \overline{v^2} = \frac{k_B T}{m}.
\end{equation}
Since the sampling is independent, $\overline{xv}=0$. The fourth order moment $\overline{x^2 v^2}$ is then given by 
\begin{equation}
    \overline{x^2v^2}=\overline{x^2}\cdot\overline{v^2}=\frac{k_B^2T^2}{km}.
\end{equation}

\subsection{Method 2}

We initialize using pure velocity kicks, i.e. $x=0$ in each realization and $v_\text{max}$ is chosen such that $\overline{v_\text{max}}=0$ and the variance 
\begin{equation}
    \overline{v_\text{max}^2} = \frac{2k_B T}{m}.
\end{equation}
Note the factor of $2$ in the variance, which is required in this method as we are always sampling the maximum velocity when the oscillator is passing through its equilibrium position ($x=0$). 

However, a thermal ensemble must have a randomized phase of oscillation, whereas in the current method the oscillator is always initialized at $x=0$ and maximum velocity. To randomize the phase, we can allow the oscillator to freely evolve for a mixing time $t_\text{mix}$ that is randomly set in each trajectory. At $t=t_\text{mix}$, the oscillator coordinates are 
\begin{equation}
    x = \frac{v_\text{max}}{\omega} \sin \omega t_\text{mix} \,\; v = v_\text{max} \cos \omega t_\text{mix},
\end{equation}
where $\omega=\sqrt{k/m}$ is the oscillator frequency. We can consider these coordinates as the new initial conditions, since we have effectively randomized the phase via $t_\text{mix}$. Clearly $\overline{x}=\overline{v}=0$, whereas 
\begin{equation}
    \overline{x^2} = \frac{\overline{v_\text{max}^2}}{\omega^2}\overline{\sin^2 (\omega t_\text{mix})} = \frac{k_B T}{k},
\end{equation}
while a similar calculation yields $\overline{v^2}=k_B T/m$. Similarly, $\overline{xv} = 0$. In computing these averages, we have used $\overline{\sin (\omega t_\text{mix})} = \overline{\cos (\omega t_\text{mix})} = 0$, $\overline{\sin^2 (\omega t_\text{mix})} = \overline{\cos^2 (\omega t_\text{mix})} = 1/2$, and $\overline{\sin (\omega t_\text{mix}) \cos (\omega t_\text{mix})} = 0$.

Let us consider the fourth moment $\overline{x^2v^2}$. This quantity evaluates to 
\begin{equation}
    \overline{x^2v^2} = \frac{\overline{v_\text{max}^4}}{\omega^2}\overline{\sin^2(\omega t_\text{mix}) \cos^2(\omega t_\text{mix})} = \frac{3}{2}\frac{k_B^2T^2}{km},
\end{equation}
where we have used $\overline{\sin^2(\omega t_\text{mix}) \cos^2(\omega t_\text{mix})}=1/8$ and $\overline{v_\text{max}^4}=3\overline{v_\text{max}^2}^2$. This last expression is valid because $v_\text{max}$ is a Gaussian random variable with zero mean.

While the second moments from both the methods agree, the fourth moments disagree. Therefore, method 2, i.e. initialization using a velocity kick followed by a random mixing time, does not strictly realize a thermal ensemble. 
 
\section{\label{app:psd}Computation of the power spectral density}

We follow a standard procedure to compute the PSD of the ion motion using tools and documentation available in MATLAB. We sketch the procedure in this Appendix. To compute the PSD, say along the $z$ direction, we first compute the Fourier transform of an ion coordinate as 
\begin{equation}
    \td{z}_j(\omega) = \frac{1}{\sqrt{\mathcal{T}_\text{tot}}} \int_0^{\mathcal{T}_\text{tot}} dt \; z_j(t) e^{-i\omega t}.
    \label{eqn:ft}
\end{equation}
Next, we compute the one-sided PSD as $P_z(\omega) = \sum_{j=1}^N \abs{\td{z}_j(\omega)}^2 +\sum_{j=1}^N \abs{\td{z}_j(-\omega)}^2$. The PSD for motion in the $x$ and $y$ directions are computed similarly. We use the coordinates in the rotating frame to eliminate the large contribution to the PSD from the rigid body crystal rotation at frequency $\omega_r$. The in-plane PSD is obtained as the sum of $P_x(\omega)$ and $P_y(\omega)$. Details of a similar procedure to compute the PSD can be found in Ref.~\cite{tang2019PoP}, where the normalization used differs from the one in Eq.~(\ref{eqn:ft}) by a factor of $\sqrt{\mathcal{T}_\text{tot}}$. Finally, to sample the thermal initial conditions, we average the PSD over $96$ realizations, or trajectories, of the crystal dynamics.

\input{paper_main.bbl}

\end{document}

%% file: paper_main.bbl
%